\documentclass[aps,prc,reprint,superscriptaddress,floatfix,showpacs,nofootinbib]{revtex4-1}
\usepackage{epsfig}
\usepackage{dcolumn}
\usepackage{bm}
\usepackage{amssymb}
\usepackage{amsmath}
\usepackage[dvipsnames]{xcolor}

\begin{document}

\title{Possible octupole deformation  of $^{208}$Pb and the ultracentral $v_2$ to $v_3$ puzzle}

\author{P. Carzon}
\email[Email: ]{pcarzon2@illinois.edu}
\affiliation{Illinois Center for Advanced Studies of the Universe \&  Department of Physics, 
University of Illinois at Urbana-Champaign, Urbana, IL 61801, USA}
\author{S. Rao}
\email[Email: ]{svr34@scarletmail.rutgers.edu }
\affiliation{Department of Physics and Astronomy, Rutgers University, Piscataway, NJ USA 08854}
\author{M. Luzum}
\email[Email: ]{mluzum@usp.br}
\affiliation{Instituto de F\'{i}sica, Universidade de S\~{a}o Paulo, C.P.
66318, 05315-970 S\~{a}o Paulo, SP, Brazil}
\author{M. Sievert}
\email[Email: ]{msievert@nmsu.edu}
\affiliation{Department of Physics, New Mexico State University, Las Cruces, NM 88003, USA}
\affiliation{Illinois Center for Advanced Studies of the Universe \&  Department of Physics, 
University of Illinois at Urbana-Champaign, Urbana, IL 61801, USA}
\author{J. Noronha-Hostler}
\email[Email: ]{jnorhos@illinois.edu}
\affiliation{Illinois Center for Advanced Studies of the Universe \&  Department of Physics, 
University of Illinois at Urbana-Champaign, Urbana, IL 61801, USA}

\date{\today}
\begin{abstract}
Recent measurements have established the sensitivity of ultracentral heavy-ion collisions to the deformation parameters of non-spherical nuclei.  In the case of ${}^{ 129}$Xe collisions, a quadrupole deformation of the nuclear profile led to an enhancement of elliptic flow in the most central collisions.  In ${}^{ 208}$Pb collisions a discrepancy exists in similar centralities, where either elliptic flow is over-predicted or triangular flow is under-predicted by hydrodynamic models; this is known as the $v_2$-to-$v_3$ puzzle in ultracentral collisions.  Motivated by low-energy nuclear structure calculations, we consider the possibility that $^{208}$Pb nuclei could have a pear shape deformation (octupole), which has the effect of increasing triangular flow in central PbPb collisions.  Using the recent data from ALICE and ATLAS, we revisit the $v_2$-to-$v_3$ puzzle in ultracentral collisions, including new constraints from recent measurements of the triangular cumulant ratio $v_3\left\{4\right\}/v_3\left\{2\right\}$ and comparing two different hydrodynamic models.  We find that, while an octupole deformation would slightly improve the ratio between $v_2$ and $v_3$, it is at the expense of a significantly worse triangular flow cumulant ratio. In fact, the latter observable prefers no octupole deformation, with $\beta_3\lesssim 0.0375$ for ${}^{ 208}$Pb, 
and is therefore consistent with the expectation for a doubly-magic nucleus even at top collider energies.
The $v_2$-to-$v_3$ puzzle remains a challenge for hydrodynamic models.
\end{abstract}

\maketitle

\section{Introduction}

The precision of heavy-ion collisions has reached the point that differences in the collective flow can be predicted between beam energies at the level of a few percent \cite{Noronha-Hostler:2015uye, Niemi:2015voa, Adam:2015ptt}.  While minor model differences remain in terms of both the initial conditions and transport coefficients, the vast majority of collaborations can now fit (and predict) the ``bread and butter" observables such as charged particle spectra and two-particle azimuthal anisotropies \cite{Song:2010mg, Bozek:2012qs, Gardim:2012yp, Bozek:2013uha, Niemi:2015qia, Ryu:2015vwa, McDonald:2016vlt, Bernhard:2016tnd, Gardim:2016nrr, Alba:2017hhe, Giacalone:2017dud, Eskola:2017bup, Weller:2017tsr, Schenke:2019ruo}.  Extractions from the above theoretical models have generally converged on a  minimum in the shear viscosity to entropy density ratio of $\eta/s\approx 0.1\pm 0.1$ with a maximum in the bulk viscosity to entropy density ratio at $\zeta/s\approx 0.15\pm0.15$.  While a number of questions still remain about the exact scale that initial conditions probe \cite{Gardim:2017ruc,Noronha-Hostler:2015coa,Gardim:2017ruc,Nagle:2018ybc,Hippert:2020kde,NunesdaSilva:2020bfs} and the limitations of hydrodynamics in small systems \cite{Acharya:2019vdf}, a general consensus has emerged within the community about the applicability of event-by-event relativistic viscous hydrodynamics in large AA collisions.  

Therefore, when there are deviations of experimental flow data from theoretical calculations in large systems, it generally signifies missing physics that goes beyond the above paradigm.  One recent example was the measurement of a significant enhancement in $v_2\{2\}$ in central $^{129}$Xe$^{129}$Xe collisions compared to expectations that assumed a spherical nucleus.  However, the $^{129}$Xe nucleus is expected to have a small quadrupole deformation, and this enhancement is explained by such an effect \cite{Giacalone:2017dud, Acharya:2018ihu, Acharya:2018eaq, CMS:2018jmx, ATLAS:2018iom}.  Previous studies \cite{Adamczyk:2015obl, Goldschmidt:2015kpa, Dasgupta:2016qkq, Giacalone:2018apa} of collisions of $^{238}$U, which is known to have a substantial quadrupole deformation, also found an enhancement in the elliptic flow in ultracentral events. In both of these cases the nuclei have a quadrupole $\beta_2$  (and a small hexadecapole $\beta_4$), leading to an enhanced elliptic shape and hence to an increase in $v_2\{2\}$ in ultracentral collisions.  Additional proposals include studying the effect of alpha clustering in $^{16}O$  \cite{Lim:2019cys,Citron:2018lsq,Rybczynski:2019adt}, scanning small deformed ions \cite{Noronha-Hostler:2019ytn}, and exploring polarized nuclear beams \cite{Bozek:2018xzy, Broniowski:2019kjo}. Recent work discussing the details of measuring deformations using heavy-ion collisions can be found in Refs.~\cite{Giacalone:2019pca, Giacalone:2020awm}. Typically the most sensitivity to nuclear deformation occurs in central collisions; this is because at zero impact parameter the anisotropy occurs solely due to fluctuations, while in other centralities the impact region is dominated by the almond overlap shape.  Furthermore, other sources of fluctuations which scale with the system size will have the smallest effect in central collisions where the overlap region is the largest. 
 
Similarly, it has been noted that ultracentral $^{208}$Pb$^{208}$Pb collisions exhibit a relationship between $v_2\{2\}$ and  $v_3\{2\}$ that is not shown by any existing theoretical model \cite{Luzum:2012wu,CMS:2012xxa,Shen:2015qta,Rose:2014fba}.   While theoretical calculations which treat the $^{208}$Pb nucleus as spherical predict a natural hierarchy of $v_2\{2\}>  v_3\{2\}$, the data instead show  $v_2\{2\} \approx { v_3\{2\} }$.  As a result, the models all either underpredict the triangular flow $v_3\{2\}$ or overpredict the elliptic flow $v_2\{2\}$ in ultracentral $^{208}$Pb$^{208}$Pb collisions.  A number of attempts have been made to resolve this puzzle by varying the initial conditions \cite{Luzum:2012wu,Shen:2015qta, Gelis:2019vzt}, the transport coefficients \cite{Luzum:2012wu,Rose:2014fba}, and the equation of state \cite{Alba:2017hhe}, but it still remains a generic feature of models.  In addition to this discrepancy among the two-particle correlations $v_2 \{2\}$ and $v_3 \{2\}$, it was also pointed out that the four-particle correlation $v_3 \{4\}$ also contains discrepancies between theory and data in very central collisions \cite{Giacalone:2017uqx}.  In the four-particle sector, the hydrodynamic models tended to underpredict the ratio $v_3\{4\}/v_3\{2\}$ \cite{Chatrchyan:2013kba, ALICE:2011ab,Aad:2014vba}, which corresponds to overestimating the width of the event-by-event fluctuations in $v_3$.

The reason that the hierarchy $v_2\{2\}>  v_3\{2\}$ emerges from hydrodynamic models in ultracentral collisions is as follows.  In heavy ion collisions and especially in central collisions of large systems, there is a strong linear mapping from the initial-state geometry to the final-state collective flow.  This mapping is quantified by the nearly linear proportionality between the initial-state eccentricity vectors $\varepsilon_n$ and the final-state flow vectors $v_n$ 
\begin{equation}
v_n=\kappa_n \varepsilon_n .
\end{equation}
Non-linear corrections to the hydrodynamic response can become significant in more peripheral collisions but are negligible in the ultracentral region considered here \cite{Noronha-Hostler:2015dbi, Sievert:2019zjr, Rao:2019vgy}.  The linear response coefficients $\kappa_n$ thus encode the entire hydrodynamic evolution and properties of the medium.

In ultracentral collisions, where the impact parameter $b$ nearly vanishes, the mean-field geometry of the collision is perfectly round (at least for spherical nuclei).  Then $\varepsilon_2 = \varepsilon_3 = 0$ on average, and it is the event-by-event fluctuations of the geometry which produce both $\varepsilon_2$ and $\varepsilon_3$, leading to $\varepsilon_2 \{2\} \approx \varepsilon_3 \{2\}$.  However, this picture is changed somewhat by the effects of viscosity.  Although both the elliptic and triangular harmonics are comparable at level of the initial eccentricities, the relative magnitudes of the final $v_2 , v_3$ depend on the size of the response coefficients $\kappa_2 , \kappa_3$, and dissipative effects naturally lead to $\kappa_3 < \kappa_2$.  Thus, initial conditions for round nuclei will generically exhibit an  $\varepsilon_2 \{2\} \approx \varepsilon_3 \{2\}$ relationship, leading to $v_2 \{2\} > v_3 \{2\}$ in the final state, in disagreement with the data.\footnote{The values of $v_2\{2\}$ and $ v_3\{2\}$  depend on the range of transverse momentum, and so one does not always have $v_2\{2\}\approx  v_3\{2\}$.  Nevertheless, the dependence is the same in theory and experiment, and so the discrepancy remains, regardless of the details of the measurement.
}  
To overcome the additional viscous suppression of $v_3$ in the final state, instead one needs an initial state with $\varepsilon_2<\varepsilon_3$.

In light of the recent studies on deformed nuclei in central heavy ion collisions, it's natural to question if $^{208}$Pb may also be showing signs of nuclear deformation.  This would be somewhat surprising  since $^{208}$Pb is a highly stable nucleus with doubly-magic atomic number; nuclear structure tables indicate that the octupole deformation, $\beta_3$, is zero \cite{Moller:2015fba}. However, a recent paper did predict a finite $\beta_3$ using a Minimization After Projection  model  \cite{Robledo:2011nf}.  Additionally, nuclear structure experiments are performed at significantly lower beam energies -- multiple orders of magnitude lower than relativistic heavy ion collisions --- and the structure of the nucleus may evolve from low to high energies. Finally, nuclear structure experiments only probe the electric charge density and its associated geometry, whereas heavy-ion collisions probe the color charge density and its associated geometry.  A difference in the geometry of electric charge from the full geometry of nuclear matter, such as a neutron skin \cite{Tarbert:2013jze}, could modify the initial geometry relevant for heavy-ion collisions.  It seems plausible then, that the deformation parameters such as $\beta_3$ could be different between the electric and color charge densities in $^{208}$Pb.  Furthermore, the neutron skin has been a point of interest for both low-energy nuclear collisions \cite{Abrahamyan:2012gp} and also for neutron stars' mass radius relationship \cite{Horowitz:2000xj,Fattoyev:2012rm,Fattoyev:2017jql}.  

In this paper we explore the possibility of an octupole deformation in $^{208}$Pb.  Since the original papers on the $v_2$-to-$v_3$ puzzle came out, the LHC has also had a subsequent run at 5.02 TeV with a higher luminosity, which has led to significantly smaller error bars in their experimental data.  Between updates in the medium effects (e.g. improvements in the equation of state) and these smaller error bars, we find that the $v_2$ to $v_3$ puzzle in ultracentral collisions is significantly smaller than originally thought.  Nevertheless, we find that an octupole deformation does not explain the remaining discrepancy.    Further, by using recent $v_3\{4\}/v_3\{2\}$ data from ATLAS \cite{Aaboud:2019sma} we can place a limit on the octupule deformation of $^{208}$Pb in our models.  The rest of this paper is structured as follows.  In Sec.~\ref{sec:IS} we discuss the properties of the initial state, including a discussion in Sec.~\ref{sec:deform} introducing the deformation parameters $\beta_n$ and the calculation in Sec.~\ref{sec:ecc} of the second and fourth cumulants of the initial-state eccentricities in two different models.  In Sec.~\ref{sec:hydro} we discuss the two hydrodynamical models we will be comparing: v-USPhydro and MUSIC.  In Sec.~\ref{sec:results} we compare the results of the two hydrodynamic simulations against data to assess the impact on the $v_2$-to-$v_3$ puzzle, and we summarize our conclusions in Sec.~\ref{sec:concl}.

\section{Initial State}
\label{sec:IS}

\subsection{Octupole deformation in $^{208}$Pb}
\label{sec:deform}

Modern simulations  of heavy-ion collisions begin by sampling the position of each nucleon in a nucleus on an event-by-event basis from a Woods-Saxon distribution, in order to obtain the geometrical structure of the heavy ion.  Generally the two-parameter Woods-Saxon distribution is given in spherical coordinates by
\begin{align} \label{e:2PF}
\rho(r,\theta) = \rho_0 \left[ 1 + \exp\left(\frac{r - R(\theta)}{a}\right)\right]^{-1} .
\end{align}
The Woods-Saxon profile \eqref{e:2PF} yields a nearly constant density for $r < R$ which transitions into an exponential falloff for $r > R$.  For a deformed nucleus, the radius $R = R(\theta)$ is not taken to be constant, but becomes a function of the angle $\theta$, which can be decomposed into a complete sum of spherical harmonics: 
\begin{align} \label{e:Rdef}
R(\theta) = R \Big(1 + \beta_2 Y_{20} (\theta) + \beta_3 Y_{30} (\theta) 
+\beta_4 Y_{40} (\theta) 
+ \cdots \Big) .
\end{align}
For a spherical nucleus we have $\beta_2=\beta_3=\beta_4=0$, with nonzero values induced by a quadrupole, octupole, or hexadecupole deformation, respectively.  In principle, the summation \eqref{e:Rdef} of $Y_{\ell m}$ extends over all $\ell \in [0,\infty)$, but for our present purposes we consider only the octupole deformation $\beta_3$.  Initial condition models will vary in the details of effects beyond the nucleon-level position sampling, such as color charge fluctuations, constitutent quarks, strings, and minijets; however, since the foundation is the Wood-Saxon distribution \eqref{e:2PF}, any deformation has a relatively generic effect.  

In Ref.~\cite{Robledo:2011nf}, a Minimization After Projection model was used to predict the octupole deformation $\beta_3$ of $^{208}$Pb, finding $\beta_3\approx 0.0375$ in the ground state.  However, using a Hartree-Fock-Bogoliubov calculation, the authors instead found $\beta_3=0$.  The authors also cite an expected $\pm 25\%$ error for the comparison of these deformation parameters between theory and experiment.  Thus, from nuclear structure considerations we expect that the octupole deformation of $^{208}$Pb should be constrained to $\beta_3 \lesssim 0.05$.  Here we check a range of $\beta_3$ for $^{208}$Pb compared to experimental data in order to quantify its effect.

In order to generate initial conditions with an octupole deformation, we have modified the Trento initial condition model \cite{Moreland:2014oya} to include a nonzero $\beta_3$.  In Trento we use the geometric mean $p=0$ (so that the entropy density at proper time $\tau = 0.6$ fm/$c$ is proportional to $\sqrt{T_A T_B}$), fluctuation parameter $k=1.6$, and nucleon width $\sigma=0.51 \, \mathrm{fm}$.  This setup was obtained from a Bayesian analysis of experimental data \cite{Bernhard:2016tnd} and has been successful in a number of predictions across systems size \cite{Giacalone:2017dud}. Additionally, we compare this to an alternative initial condition satisfying a linear scaling $T_R \propto T_A T_B$ due to the recent discussion in Refs.~\cite{Lappi:2006hq, Nagle:2018ybc, Romatschke:2017ejr, Chen:2015wia}, which we have also modified Trento to include. With linear scaling we must also adjust the multiplicity fluctuations and there we considered $k=20$.   However, all hydrodynamic simulations are performed using $p=0 $ Trento.

\subsection{Eccentricities}
\label{sec:ecc}

\begin{figure}[ht]
\includegraphics[width=0.5\textwidth]{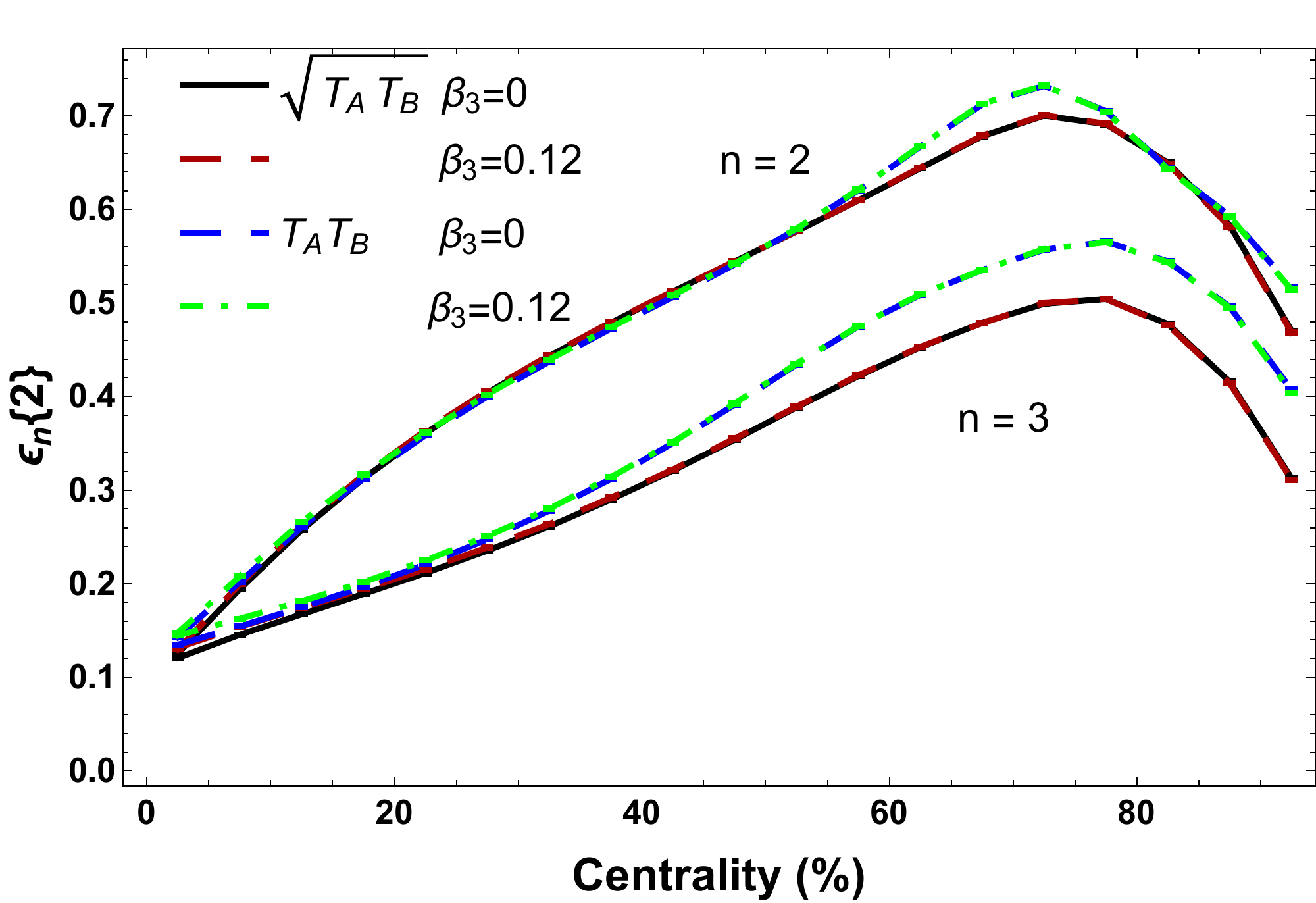} \\
\caption{(Color online) The RMS eccentricities $\varepsilon_n\{2\} = \sqrt{\langle \varepsilon_n^2 \rangle}$ for the elliptic $n=2$ and triangular $n=3$ harmonics as a function of collision centrality for $\sqrt{s_{NN}}=5.02$ TeV Pb Pb collisions.  Different curves reflect the choices of initial entropy deposition ($\sqrt{T_A T_B}$ vs $T_A T_B$) and octupole deformation $\beta_3$. 
}
\label{fig:ecc2}
\end{figure}

While the impact parameter in heavy ion collisions is not directly measurable, proxy measures such as the number of produced particles (which is anticorrelated with impact parameter) can be used instead.  When binned into centrality percentiles, the produced multiplicity is greatest in the events with complete overlap of the ions and small impact parameter.  For the most central collisions (say $0-1\%$ centrality), the impact parameter is essentially zero; finer binning into ultra-central collisions is additionally sensitive to nuclear deformation and to fluctuations in particle production \cite{Noronha-Hostler:2019ytn}.  

Since the number of produced particles in the final state is controlled by the entropy of the system at freeze-out, centrality binning by entropy is a good approximation to centrality determined by multiplicity.  Moreover, due to the low viscosity of the quark-gluon plasma, the total entropy is nearly conserved, allowing us to approximately determine an event's centrality directly from the initial state. Then, in a given event, the eccentricities calculated in the center of mass frame are defined as 
\begin{equation}    \label{eq:vnepsn}
\varepsilon_n= - \frac{\int r^n e^{in\phi}s(r,\phi) r drd\phi}{\int r^n s(r,\phi) r drd\phi}
\end{equation}
with $s(r, \phi)$ the entropy density in polar coordinates.  In the center of mass frame, $\varepsilon_1 = 0$ identically\footnote{Not to be confused with the dipole asymmetry, which is often notated as $\varepsilon_1$}, and $\varepsilon_2, \varepsilon_3$ quantify the elliptic and triangular shape of the impact region, respectively.  Because of the (anti)correlation between impact parameter and centrality, there is a strong dependence of the ellipticity on centrality:  $\varepsilon_2$ is maximal in mid-central collisions where the elliptical shape of the overlap region is most prominent.

Compared to the ellipticity $\varepsilon_2$, the triangularity $\varepsilon_3$ is zero at a mean-field level, being driven entirely by the event-by-event fluctuations away from a smooth Woods-Saxon profile \eqref{e:2PF}.  As a result, the dependence of $\varepsilon_3$ versus centrality is therefore flatter than for $\varepsilon_2$, growing only as the transverse size of the collision area decreases, as shown in Fig.\ \ref{fig:ecc2}. In fact, one can vary $\beta_3$ quite significantly without any visible effect in mid-central to peripheral collisions.  Only in central collisions (i.e. $0-10\%$) are differences visible for a deformed nucleus with nonzero $\beta_3$.

\begin{figure*}[ht]
\begin{tabular}{c c}
\includegraphics[width=0.5\textwidth]{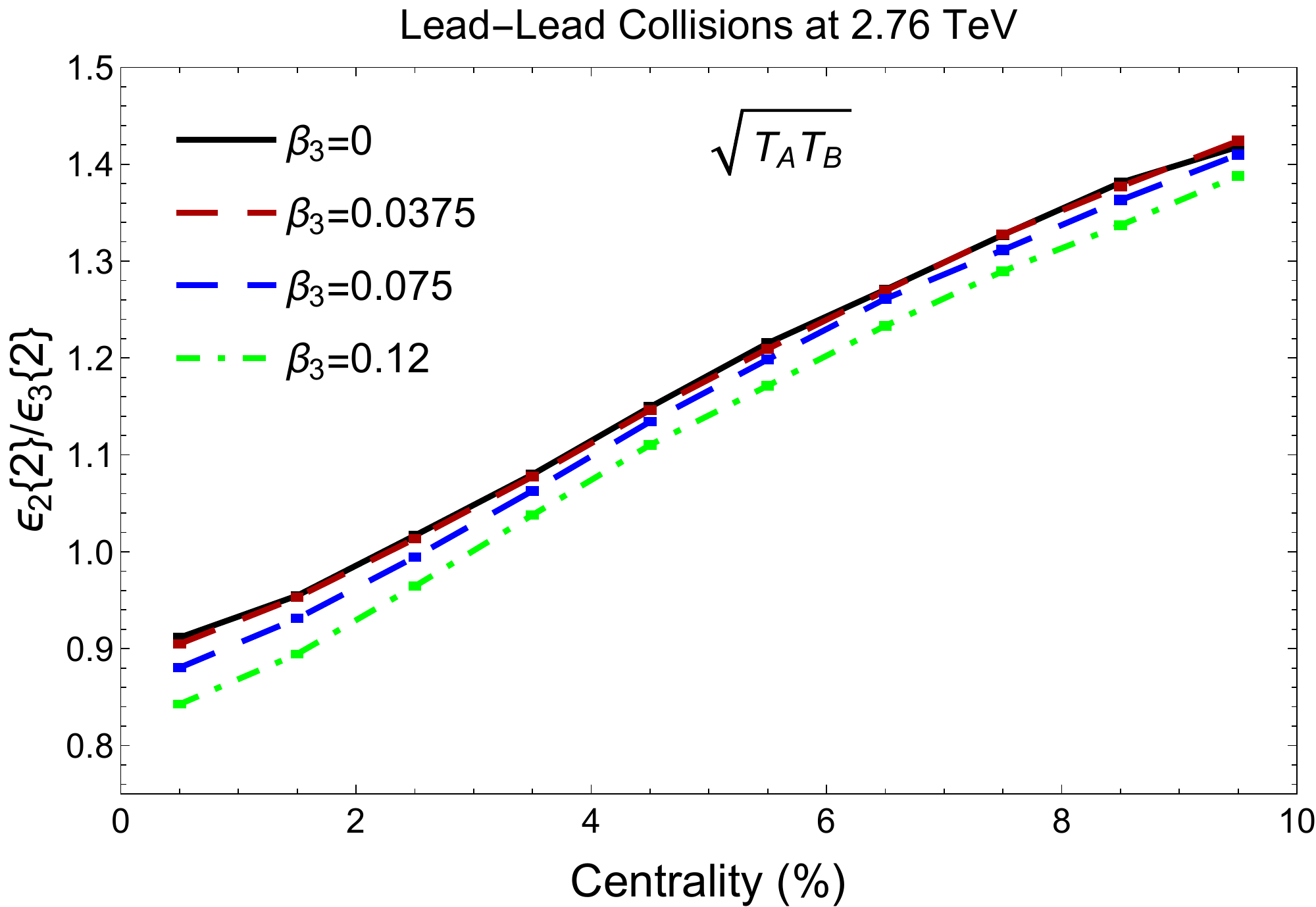} &\includegraphics[width=0.5\textwidth]{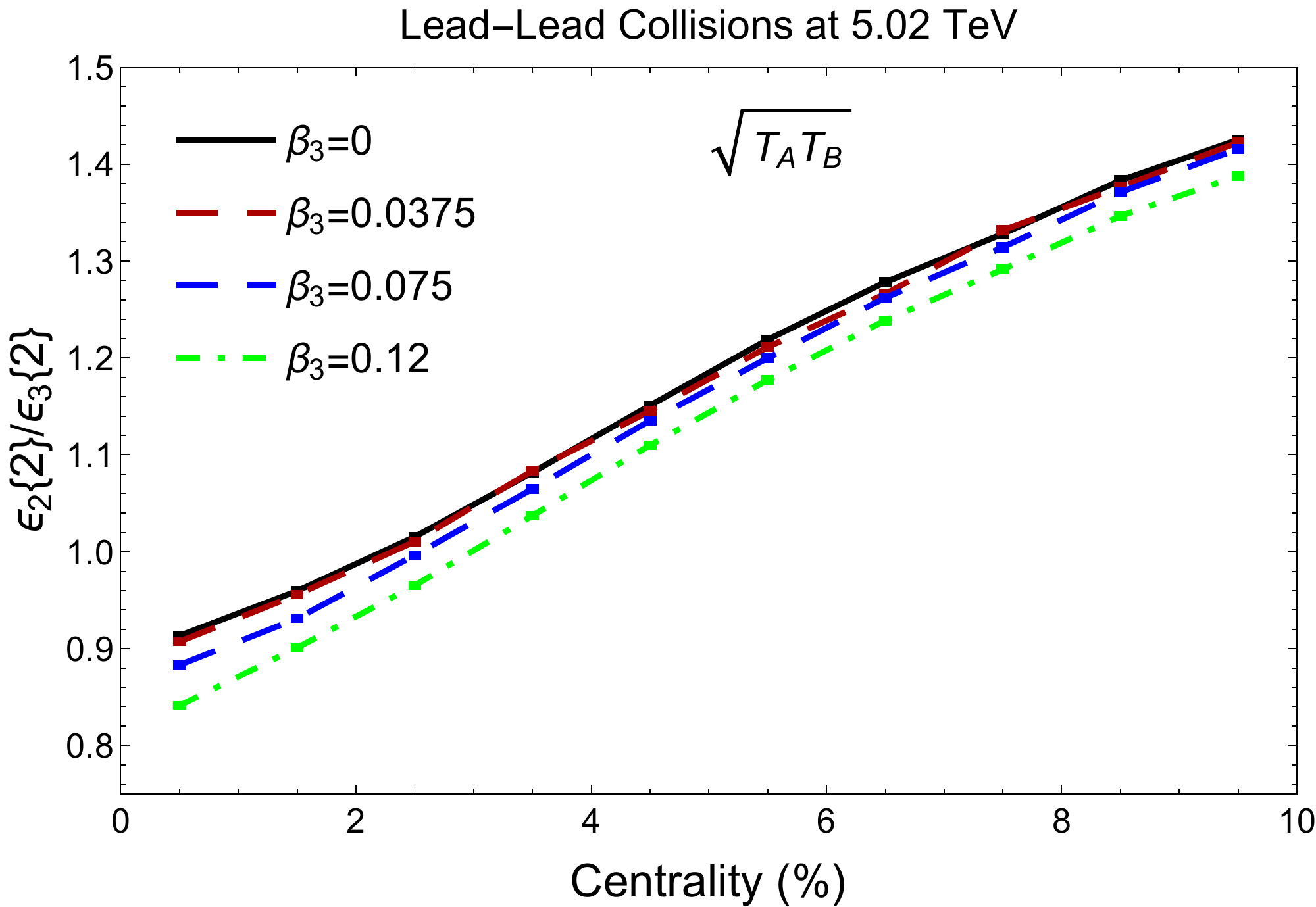} \\
\includegraphics[width=0.5\textwidth]{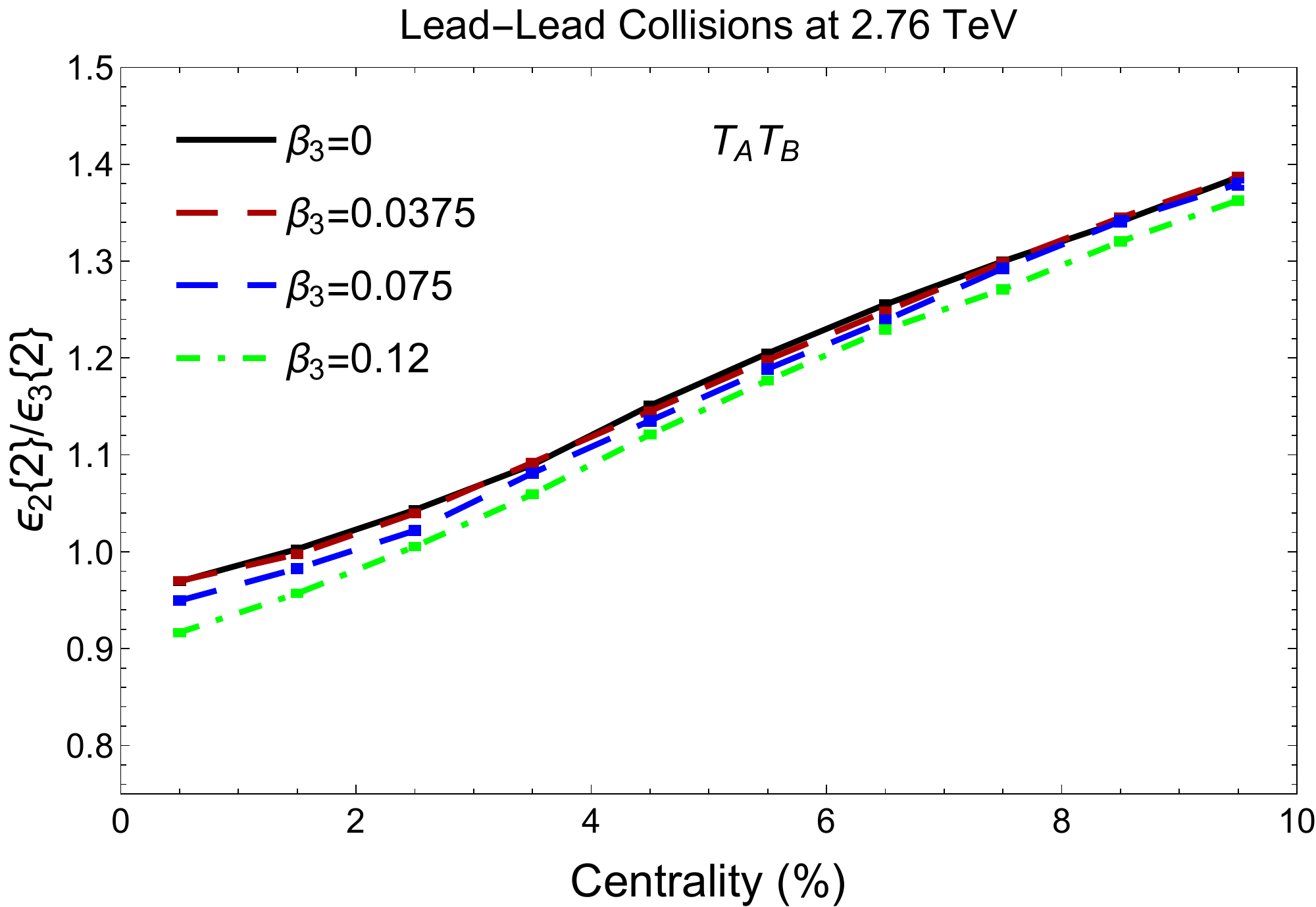} & \includegraphics[width=0.5\textwidth]{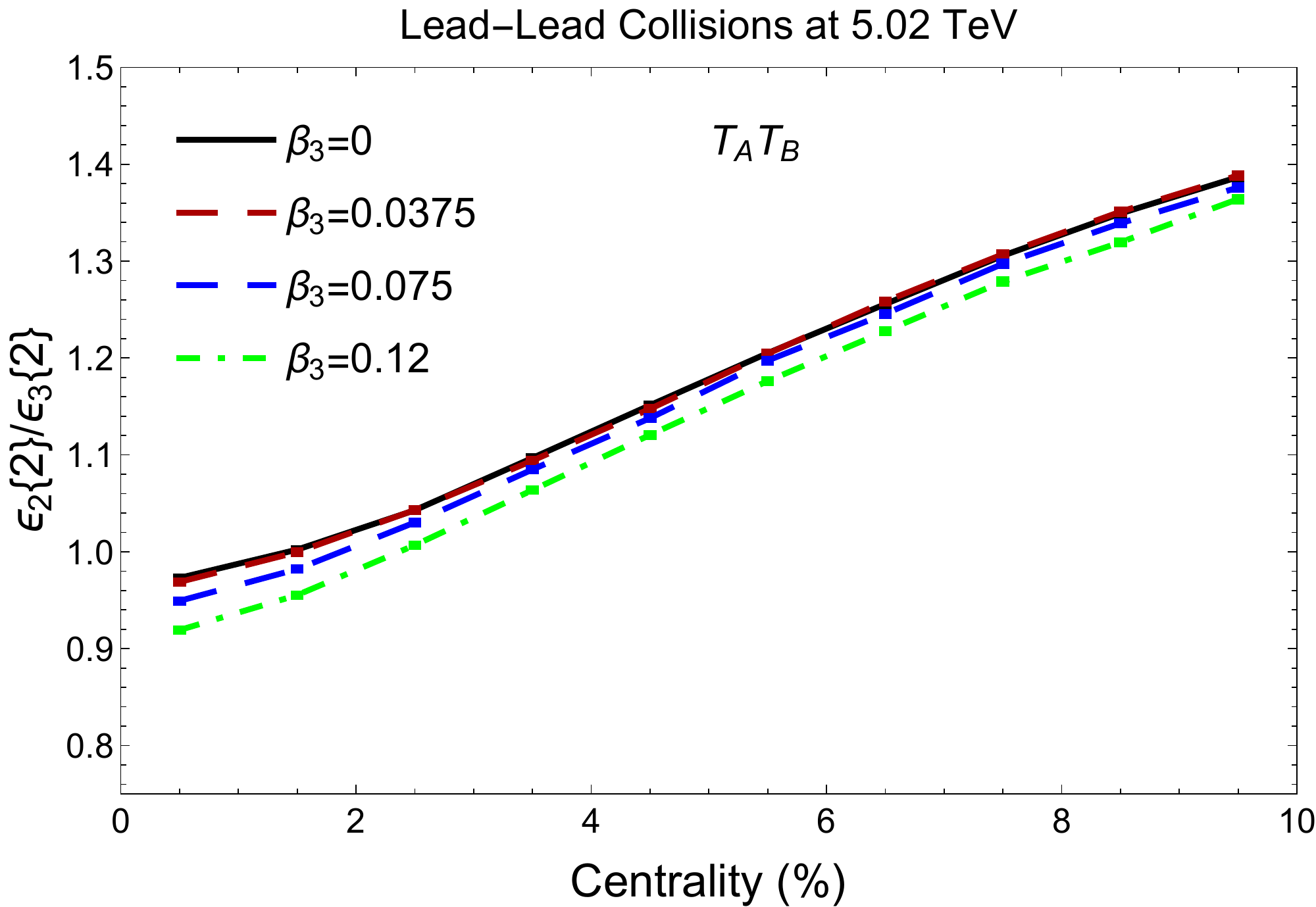} 
\end{tabular}
\caption{(Color online) Ratio of $\varepsilon_2\{2\}/\varepsilon_3\{2\}$ for PbPb collisions at $2.76$ TeV (left) and $5.02$ TeV (right) for varying ocutpole deformations $\beta_3$. We also compare two models for the initial-state entropy deposition: $s \approx \sqrt{T_AT_B}$ (top row) and $s \approx T_A T_B$ (bottom row).  
}
\label{fig:rat}
\end{figure*}

In Fig.\ \ref{fig:rat} we study the central collisions in more detail by comparing the ratio $\frac{\varepsilon_3 \{2\}}{\varepsilon_2 \{2\}}$ of the triangularity to ellipticity.  This initial-state quantity is not equal to $\frac{v_3 \{2\}}{v_2 \{2\}}$ in the final state because the response coefficients $\kappa_n$ in Eq.~\eqref {eq:vnepsn} do not cancel; still a change in eccentricity ratio will be reflected directly in the final-state flow ratio. We show results for Pb Pb collisions at both 2.76 TeV and 5.02 TeV for a range of $\beta_3$ octupole values.  We also compare the eccentricities obtained from the Bayesian-preferred geometric mean $s \approx \sqrt{T_A T_B}$ of Trento versus those of a linear scaling $s \approx T_A T_B$, inspired by CGC expectations for the mean energy density at early times \cite{Lappi:2006hq, Nagle:2018ybc, Romatschke:2017ejr, Chen:2015wia}.  We find that $\sqrt{T_AT_B}$ is slightly more likely to provide eccentricities compatible with experimental data. This is because in the most central collisions that we consider here ($0-1\%$) we find that $\sqrt{T_A T_B}$ leads to $\varepsilon _3\left\{2\right\} > \varepsilon _2\left\{2\right\}$ by $10-20\%$, compared to the linear $T_A T_B$ scaling which leads to a smaller enhancement of only a few percent.  As discussed previously, initial conditions with $\varepsilon_3 > \varepsilon_2$ in central collisions are needed in order to obtain the observed $v_3 \approx v_2$ after viscous damping of the higher harmonics.  Finally, we note that there does not appear to be a strong beam energy dependence to this ratio, consistent with other studies \cite{Rao:2019vgy}.

\begin{figure*}[ht]
\begin{tabular}{c c}
\includegraphics[width=0.5\textwidth]{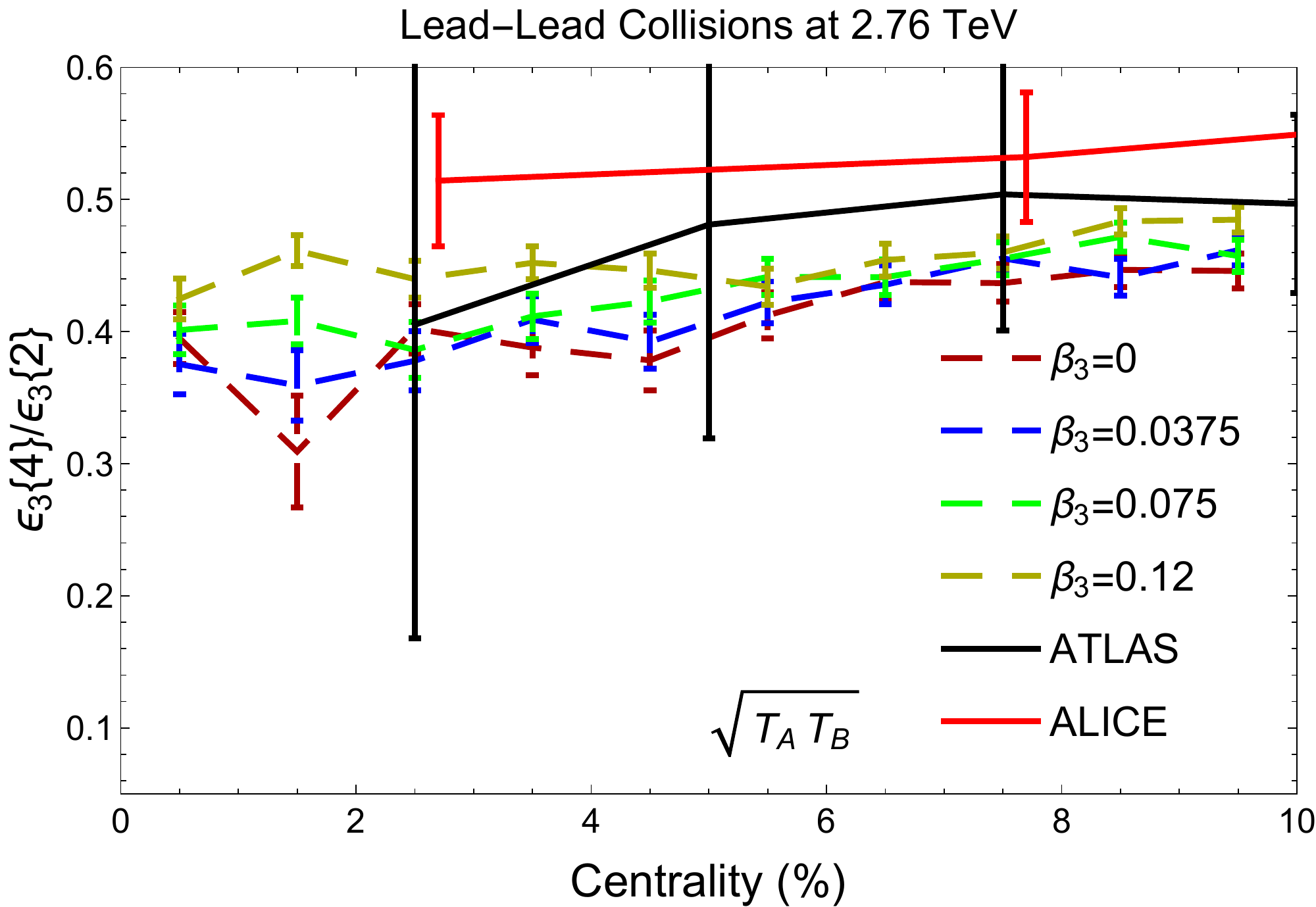}
&\includegraphics[width=0.5\textwidth]{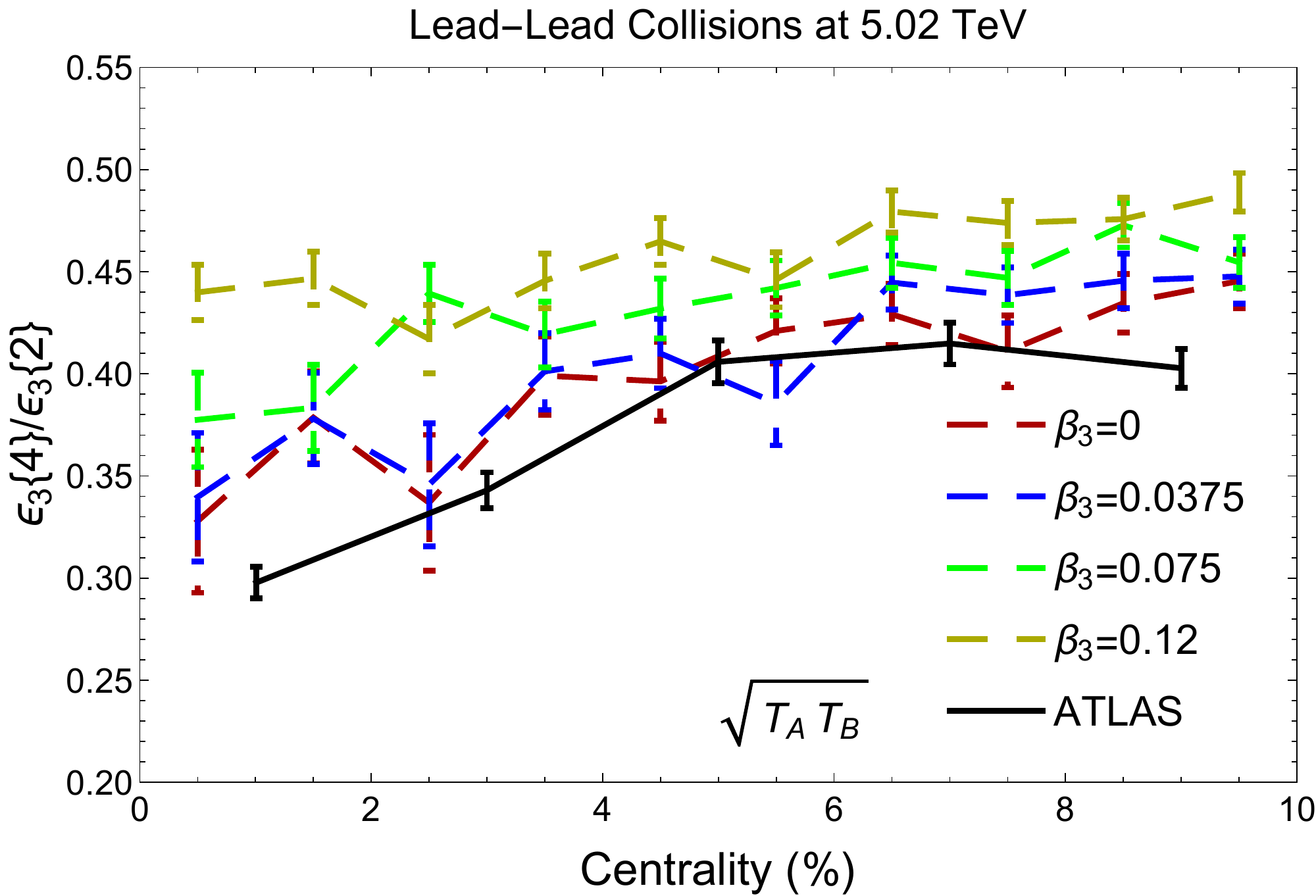}
\end{tabular}
\begin{tabular}{c c}
\includegraphics[width=0.5\textwidth]{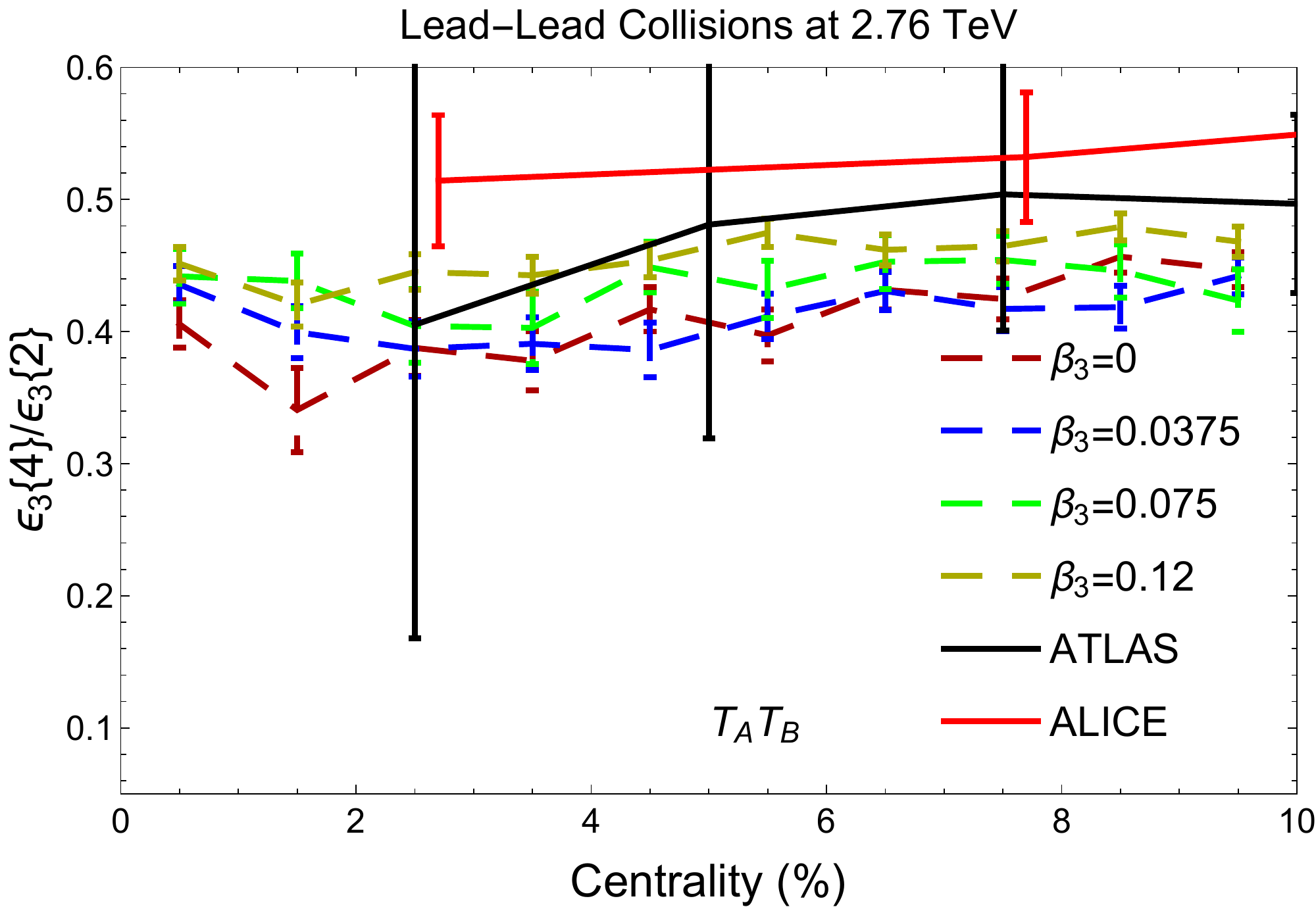} &\includegraphics[width=0.5\textwidth]{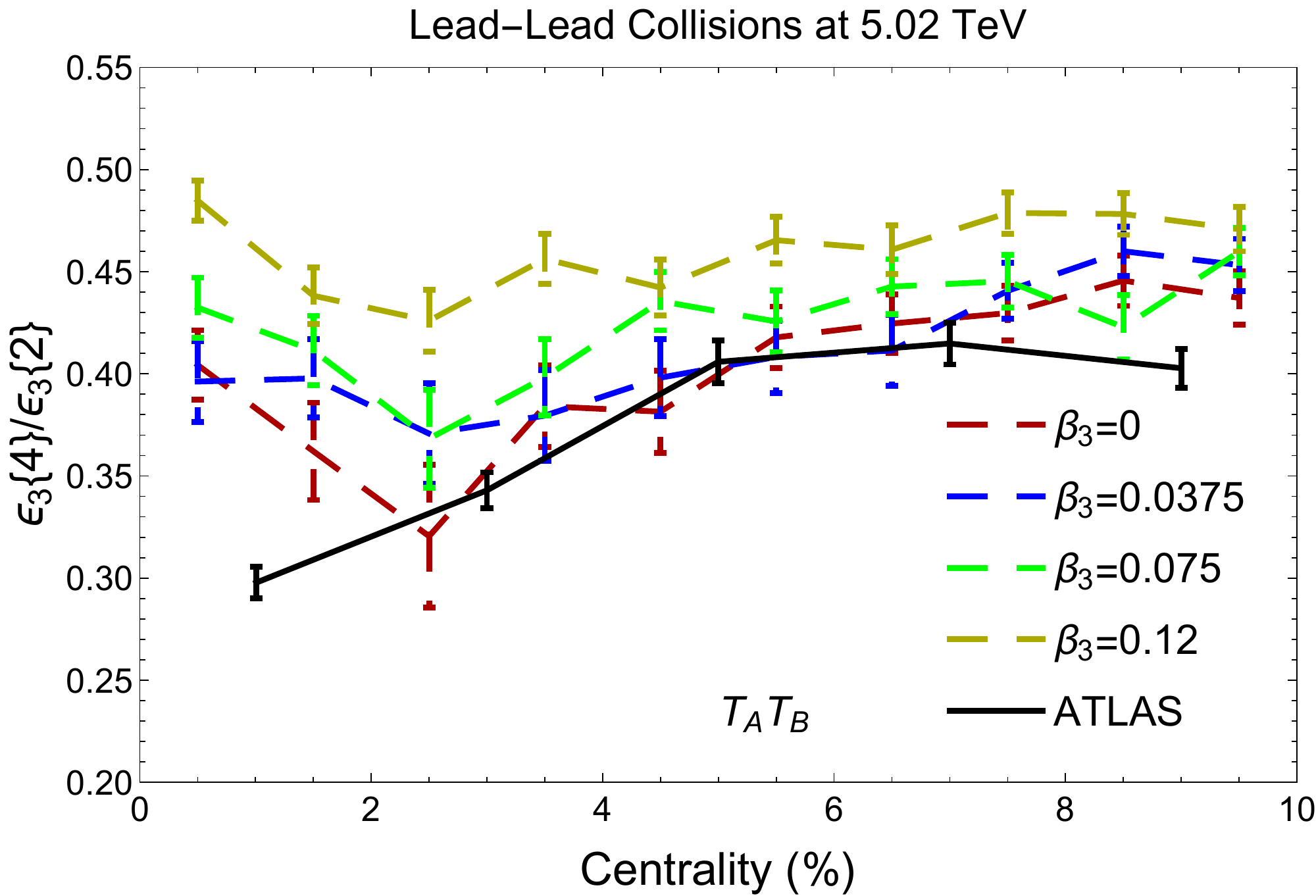} 
\end{tabular}
\caption{(Color online) Ratio of $\varepsilon_3\{4\}/\varepsilon_3\{2\}$ for Pb Pb collisions at $2.76$ TeV (left) and $5.02$ TeV (right) for varying octupole deformations $\beta_3$. We also compare two models for the initial-state entropy deposition: $s \approx \sqrt{T_AT_B}$ (top row) and $s \approx T_A T_B$ (bottom row).  We compare against data at 2.76 TeV from ALICE \cite{ALICE:2011ab} and ATLAS \cite{Aad:2014vba} and at 5.02 TeV from ALICE \cite{Acharya:2018lmh, Acharya:2019vdf}.
}
\label{fig:v34v32}
\end{figure*}

While the ratio of $v_2\{2\}$ and $v_3\{2\}$ has received the most attention, there are also important discrepancies in the ratio of $v_3\{4\}/v_3\{2\}$ predicted by models.  In Ref.~\cite{Giacalone:2017uqx} it was pointed out that 
\begin{equation}
\frac{v_3\{4\}}{v_3\{2\}}\approx\frac{\varepsilon_3\{4\}}{\varepsilon_3\{2\}}
\end{equation}
due to the nearly-linear response in central collisions.  To ensure that this ratio holds, we ran 11,000 events in the $0-1\%$ centrality class and compared the initial-state ratio of eccentricities versus the final-state hydrodynamic flow.  We found that the difference between $\frac{v_3\{4\}}{v_3\{2\}}$ and $\frac{\varepsilon_3\{4\}}{\varepsilon_3\{2\}}$ was less than $1\%$, giving confidence that we can compare the eccentricity ratio for the initial condition models directly.  This assumption is important since this observable requires extremely high statistics, on the order of 15 million events to cover a wide centrality range, so that running full hydrodynamic simulations would be prohibitive.  

In Fig.\ \ref{fig:v34v32} we plot the cumulant ratio $\frac{\varepsilon_3\{4\}}{\varepsilon_3\{2\}}$ as $\beta_3$ is varied and find that the larger $\beta_3$, the more this ratio is enhanced.  The previous comparisons for this observable \cite{Giacalone:2017uqx} were for Pb Pb collisions at 2.76 TeV where the estimated experimental errors were quite large.  Importantly, these uncertainties are certainly overestimated because the covariance was not reported by the experimental collaborations, so that the uncertainties in the ratio overcount the correlated errors.  Comparing the $\beta_3$ variations to previous experimental results at 2.76 TeV, we find that no range of $\beta_3$ can be eliminated from $\frac{\varepsilon_3\{4\}}{\varepsilon_3\{2\}}$ and, in fact, $\beta_3=0.12$ appears to be preferred compared to ALICE results \cite{Acharya:2019vdf}. 

However, due to the higher luminosity at 5.02 TeV and because the ATLAS experiment did report the complete uncertainty for the ratio $\frac{v_3\{4\}}{v_3\{2\}}$, this ratio can be used to put constraints on the $\beta_3$ parameter.  Since in this ratio the linear response coefficient $\kappa_3$ cancels, it is mostly sensitive to the choice of initial state models rather than medium properties.  In fact, for $\sqrt{T_AT_B}$ scaling only $\beta_3=0$ and $\beta_3=0.0375$ fit within the ATLAS experimental error bands, which allows us to set the upper bound $\beta_3 \leq 0.0375$. Interestingly, $T_AT_B$ scaling consistently overpredicts ATLAS data and appears to be significantly less sensitive to the $\beta_3$ deformation parameter compared to $\sqrt{T_AT_B}$.  While $\sqrt{T_AT_B}$ scaling is not well-motivated from a theoretical perspective, it appears to better capture the features of the experimental data. 

\begin{figure}[ht]
\includegraphics[width=0.48\textwidth]{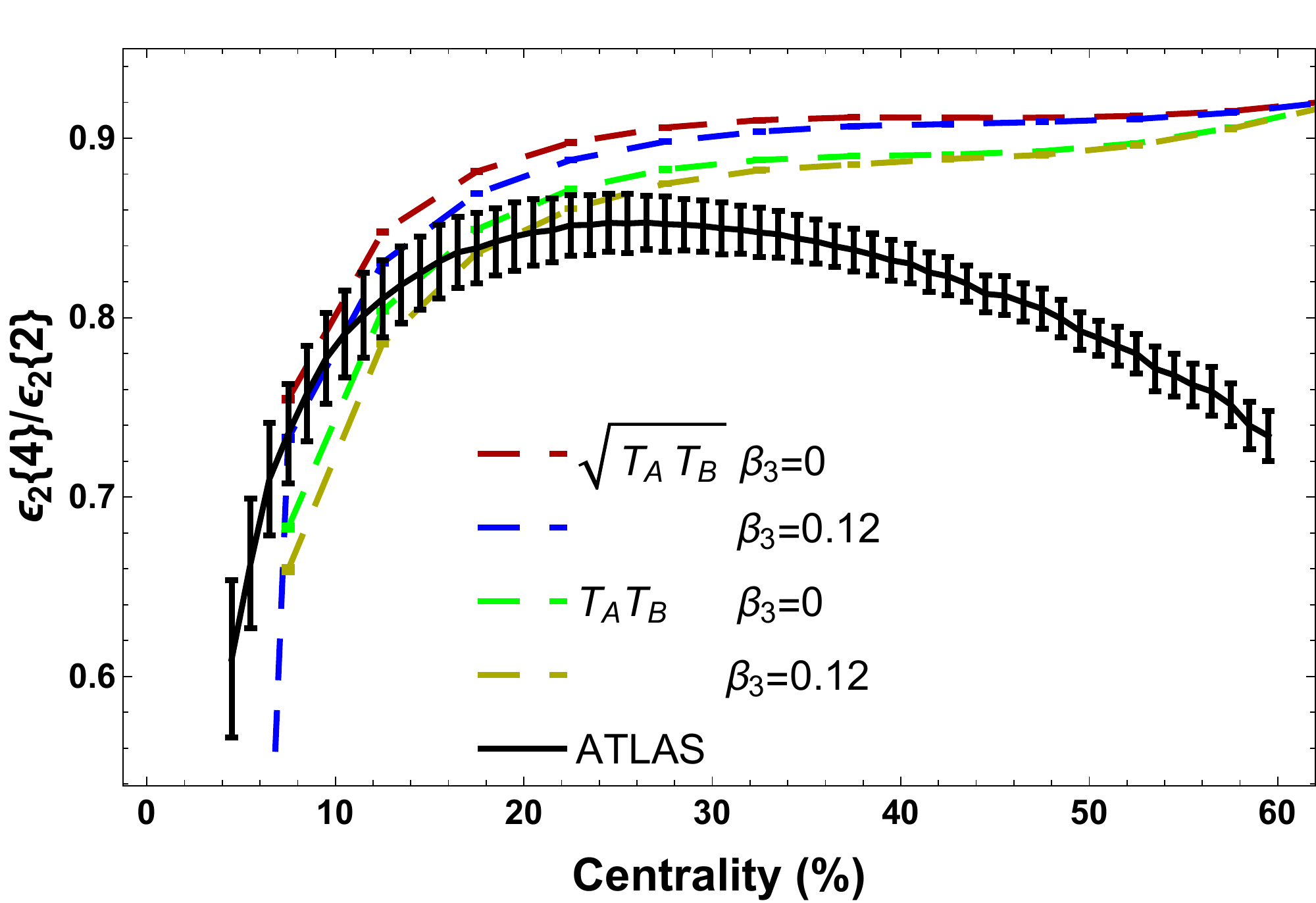}\\
\caption{(Color online) Ratio of $\varepsilon_2\{4\}/\varepsilon_2\{2\}$ for PbPb collisions at $5.02$ TeV for varying octupole deformations, $\beta_3$, and $\sqrt{T_AT_B}$ and $T_AT_B$ scaling.  We compare against data at 2.76 TeV from ATLAS \cite{Aad:2014vba}.}
\label{fig:e24e22}
\end{figure}

For completeness, in Fig.\ \ref{fig:e24e22} we also plot the cumulant ratio $\frac{\varepsilon_2\{4\}}{\varepsilon_2\{2\}}$ for $\beta_3$ of 0 and 0.12 for both scalings. From 0-20\% Centrality there is good agreement to ALICE data \cite{Acharya:2019vdf} from the $\beta_3=0$ deformations with the higher deformation $\beta_3=0.12$ showing some tension with data in very central collisions. Above 30\% Centrality all models start to deviate significantly from data, reflecting the onset of nonlinear response in mid-central collisions. 

\section{Hydrodynamical model}
\label{sec:hydro}

We study the hydrodynamic response to these initial conditions by performing full simulations, and to ensure the results are not unique to a particular hydrodynamic model, we perform calculations within two different frameworks for comparison.  

The first model uses the hydrodynamic code v-USPhydro \cite{Noronha-Hostler:2013gga,Noronha-Hostler:2014dqa} (Parameter 1), which solves the equations of motion using Smoothed Particle Hydrodynamics.  There 
the shear viscosity to entropy density ratio is taken to be a constant $\eta/s=0.047$, the bulk viscosity is set to $\zeta/s=0$, and the freeze-out temperature is $T_{FO}=150$ MeV.  The equation of state used is WB21/PDG16+ \cite{Alba:2017hhe}, which has been shown before to fit experimental data well across system sizes and beam energies \cite{Alba:2017hhe,Giacalone:2017dud}  with other predictions in ArAr and OO systems still remaining to be confirmed \cite{Sievert:2019zjr}. One crucial emphasis of this model is on the detailed treatment of feed-down decays: the hydrodynamic phase is followed by direct decays \cite{Sollfrank:1991xm,Wiedemann:1996ig} using all known resonances from the PDG16+ \cite{Alba:2017mqu}.  The philosophy behind this method is to use the most state-of-the-art list of particles, which most closely matches Lattice QCD results \cite{Alba:2017mqu} but this sacrifices hadronic transport afterwards because many of these resonances decay into 3 or 4 bodies, which have not yet been incorporated into SMASH \cite{Ono:2019ndq} or URQMD. 

\begin{figure}[ht]
	\includegraphics[width=\linewidth]{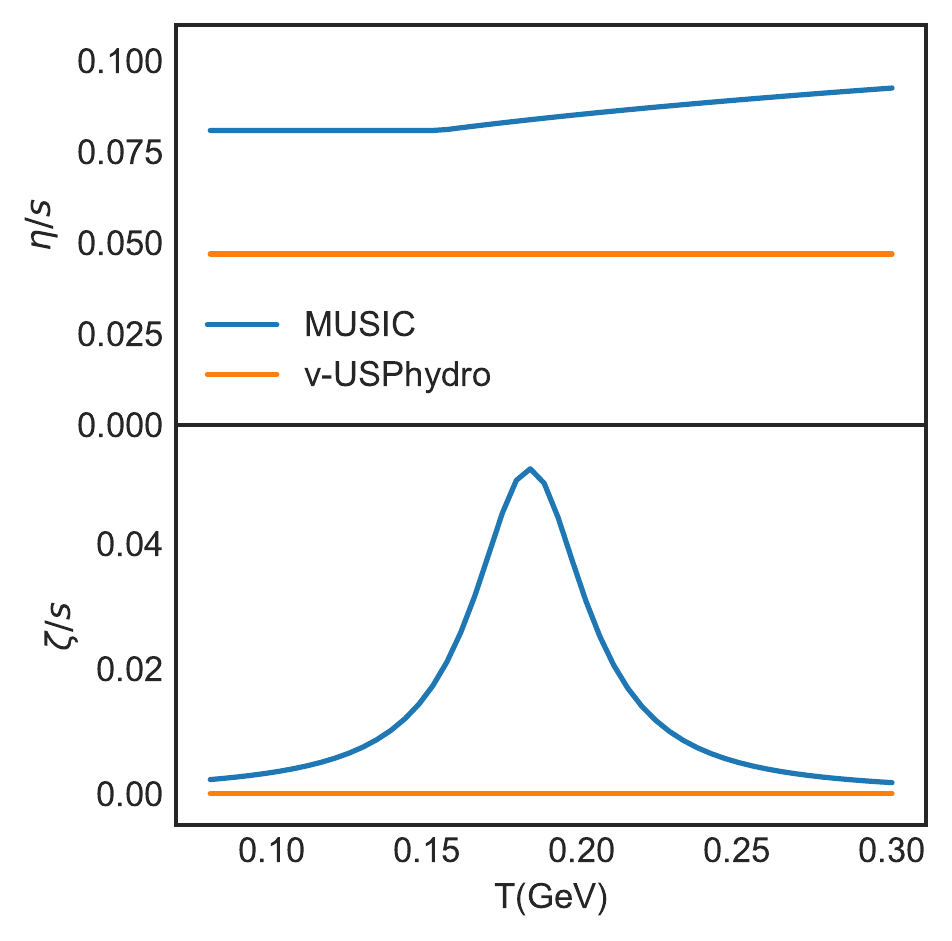} \\
	\caption{
		\label{fig:visc}
		Comparison of the bulk and shear viscosities entering the two hydrodynamic models using MUSIC and v-USPhydro.  }
	\label{f:viscosities}
\end{figure}

The second hydrodynamic model uses  MUSIC \cite{Schenke:2010nt,Schenke:2010rr} (Parameter 2), a grid-based code.   Here all fluid parameters are taken as the Maximum a Posteriori (MAP) values from the Bayesian analysis of momentum integrated observables reported in Table 5.9 of  Ref.~\cite{Bernhard:2016tnd}.  
Both $\eta/s(T)$ and $\zeta/s(T)$ depend on temperature,  as shown in Fig.~\ref{fig:visc}.  The equation of state used is \textit{s95p-v1.2 }\cite{Huovinen:2009yb}.  When the system reaches a local temperature of 151 MeV, the fluid is converted to hadrons and resonances, which evolve in the hadron cascade code UrQMD  \cite{Bass:1998ca, Bleicher:1999xi}.  These parameters were tuned for a Trento initial condition of energy density at $\tau=0$ followed by a period of free streaming $\tau_{\rm fs}=1.16$ fm/$c$, whereas here we use Trento to initialize the entropy density at a finite time $\tau_0 = 0.6$ fm/$c$ with no initial transverse flow.  Nevertheless, the fit to data at most centralities is expected to be reasonable \cite{NunesdaSilva:2018viu, NunesdaSilva:2020bfs}.

In both cases, v-USPhydro and MUSIC pass analytical solutions (the Gubser test) \cite{Marrochio:2013wla} so that any differences in the results should lie not in numerics, but in the different physical ingredients such as the shear and bulk viscosities, the equation of state, the list of hadrons and resonances, and hadronic rescatterings. Note that the hydrodynamic equations of motion are different  for v-USPhydro and MUSIC.  v-USPhydro uses phenomenological Israel-Stewart and MUSIC used DNMR with the second order transport coefficients from \cite{Denicol:2014vaa}.

\section{Results} 
\label{sec:results}

We ran our full hydrodynamic machinery for 5,000-10,000 events in the centrality range of $0-10\%$ for PbPb 5.02 TeV collisions at the LHC.  Because these simulations take a significant amount of time to run, we only consider the geometric mean scenario $T_R = \sqrt{T_A T_B}$ ($p=0$), which we found in the previous section is more likely to fit experimental data. 

\begin{figure}[ht]
\includegraphics[width=0.5\textwidth]{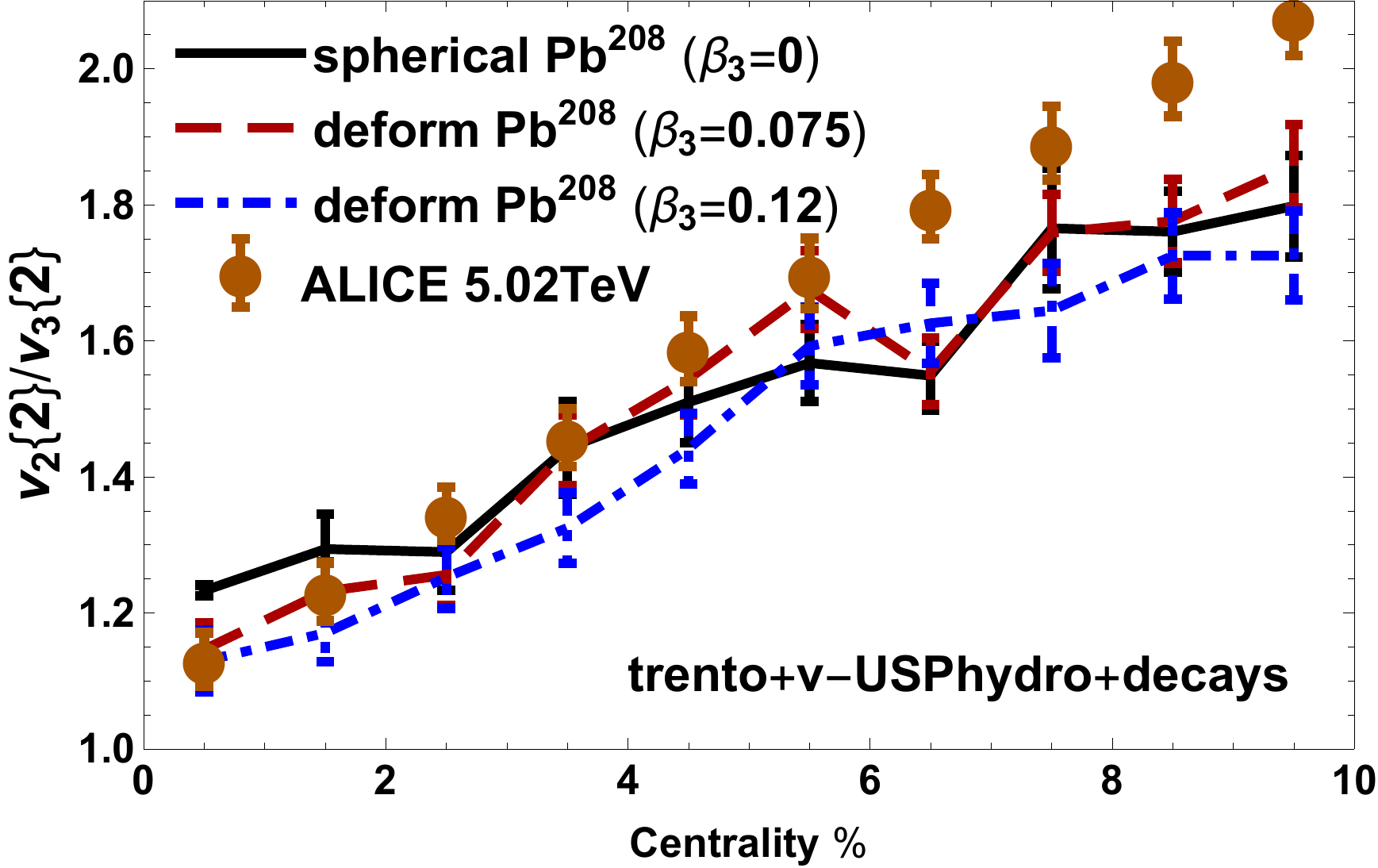} 
 \includegraphics[width=0.5\textwidth]{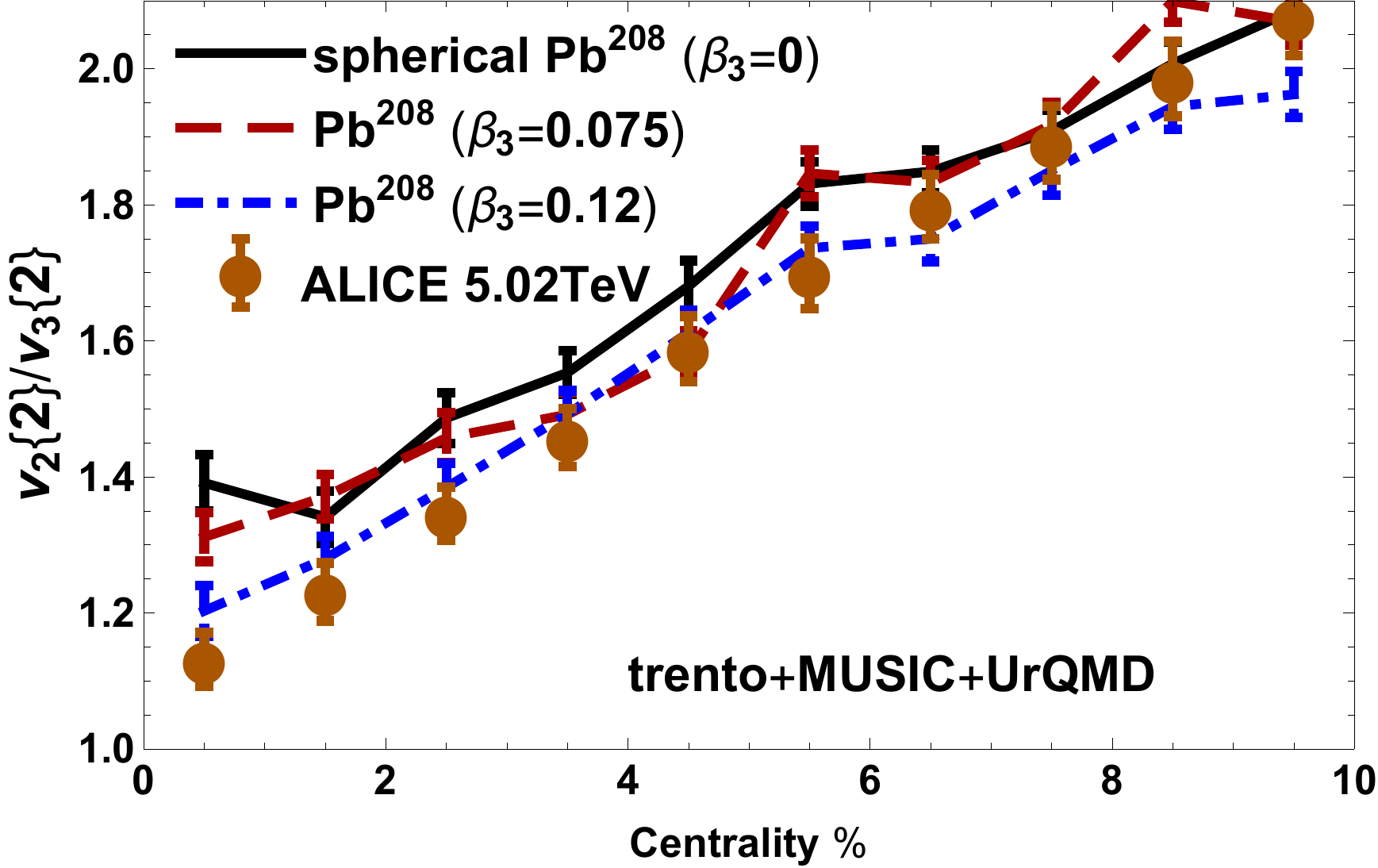} 
\caption{(Color online)  Ratio of $v_2\{2\}/v_3\{2\}$ in PbPb 5.02TeV collisions compared to experimental data from ALICE \cite{Acharya:2018lmh, Acharya:2019vdf}.}
\label{fig:v2/v3}
\end{figure}

While CMS produced ultracentral data in LHC Run 1, ALICE has significantly more data for Run 2 \cite{Acharya:2019vdf}, so here we make the comparisons for the LHC Run 2 using $0.2 \, \mathrm{GeV} < p_T < 3 \, \mathrm{GeV}$ \cite{Acharya:2018lmh}.  In Fig.\ \ref{fig:v2/v3} we plot the ratio $v_2 \{2\} \, / \, v_3 \{2\}$ for Trento+v-USPhydro+decays (top panel) and Trento+MUSIC+UrQMD (bottom) along with the ALICE data.  We see that, in both cases a finite $\beta_3$ deformation significantly improves the fit to experimental data.  For Trento+v-USPhydro+decays $\beta_3=0.075$ is the best fit, whereas for Trento+MUSIC+UrQMD $\beta_3=0.12$ provides a better fit.  We also note that the absolute magnitudes of $\frac{v_2\{2\}}{v_3\{2\}}$ are smaller in v-USPhydro than in MUSIC, which can be attributed to the smaller shear viscosity $\eta/s$ used in v-USPhydro.  This is because, unlike the ratio $\frac{v_3 \{4\}}{v_3 \{2\}}$ of different cumulants of the same harmonic, the ratio $\frac{v_2\{2\}}{v_3\{2\}}$ has significant a viscosity dependence; $v_3\{2\}$ is more strongly damped by viscosity than $v_2\{2\}$.

\begin{figure*}[t]
 \includegraphics[width=\linewidth]{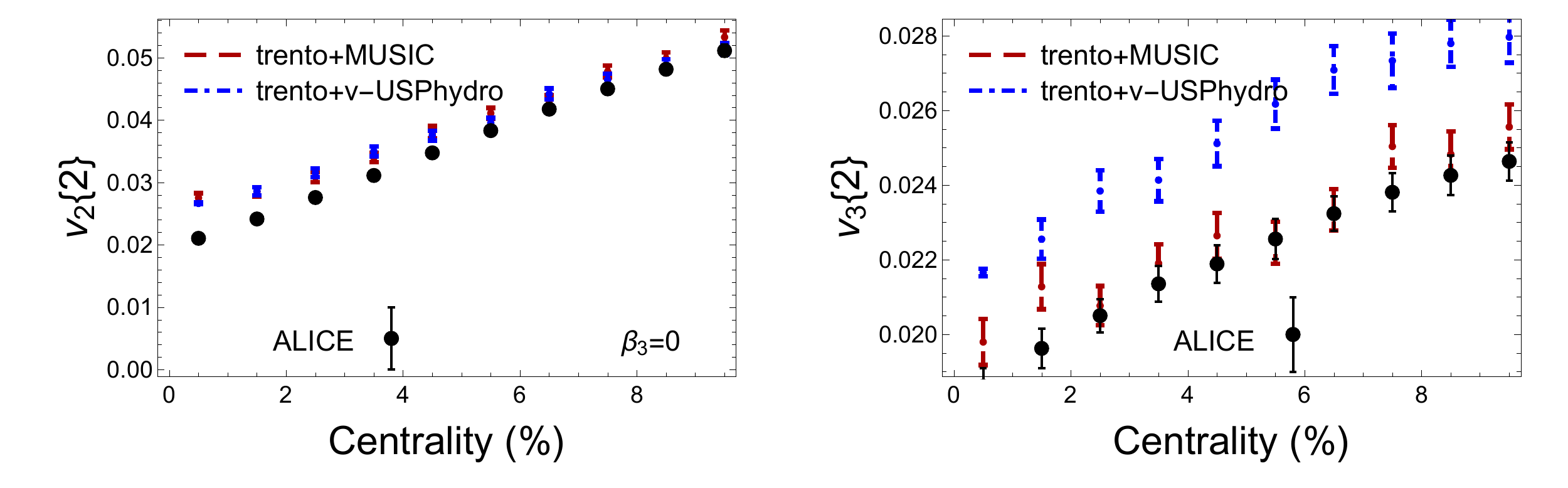} 
\caption{(Color online)  Comparison of $v_2\{2\}$ and $v_3\{2\}$ in PbPb 5.02TeV collisions from v-USPhydro and MUSIC compared to experimental data from ALICE \cite{Acharya:2018lmh, Acharya:2019vdf}.}
\label{fig:vnabs}
\end{figure*}

Comparisons of the ratio $v_2 \{2\} \, / \, v_3 \{2\}$ shown in Fig.\ \ref{fig:v2/v3} are one important element of the $v_2$-to-$v_3$ puzzle, but it is also essential to compare the absolute magnitudes of both quantities as well.  These comparisons are shown in Fig.\ \ref{fig:vnabs} for the spherical case $\beta_3 = 0$.  Generally, both Trento+MUSIC and Trento+v-USPhydro fit $v_2\left\{2\right\}$ well for centralities larger than $10\%$ centrality, but in the central collisions shown here, they both overpredict the experimental data.  Meanwhile, Trento+MUSIC manages to fit $v_3\left\{2\right\}$ quite well from $0-10\%$ centrality while Trento+v-USPhydro overpredicts the data.  Thus, based on these two representative examples of hydrodynamic models, it appears that the crux of the $v_2$-to-$v_3$ puzzle may lie in the overprediction of $v_2\left\{2\right\}$ rather than an underprediction of $v_3\left\{2\right\}$.  From this perspective, it seems unlikely that boosting the triangularity $\varepsilon_3$ by adding an octupole deformation $\beta_3$ will improve the agreement with data.

\begin{figure}[h]
 \includegraphics[width=\linewidth]{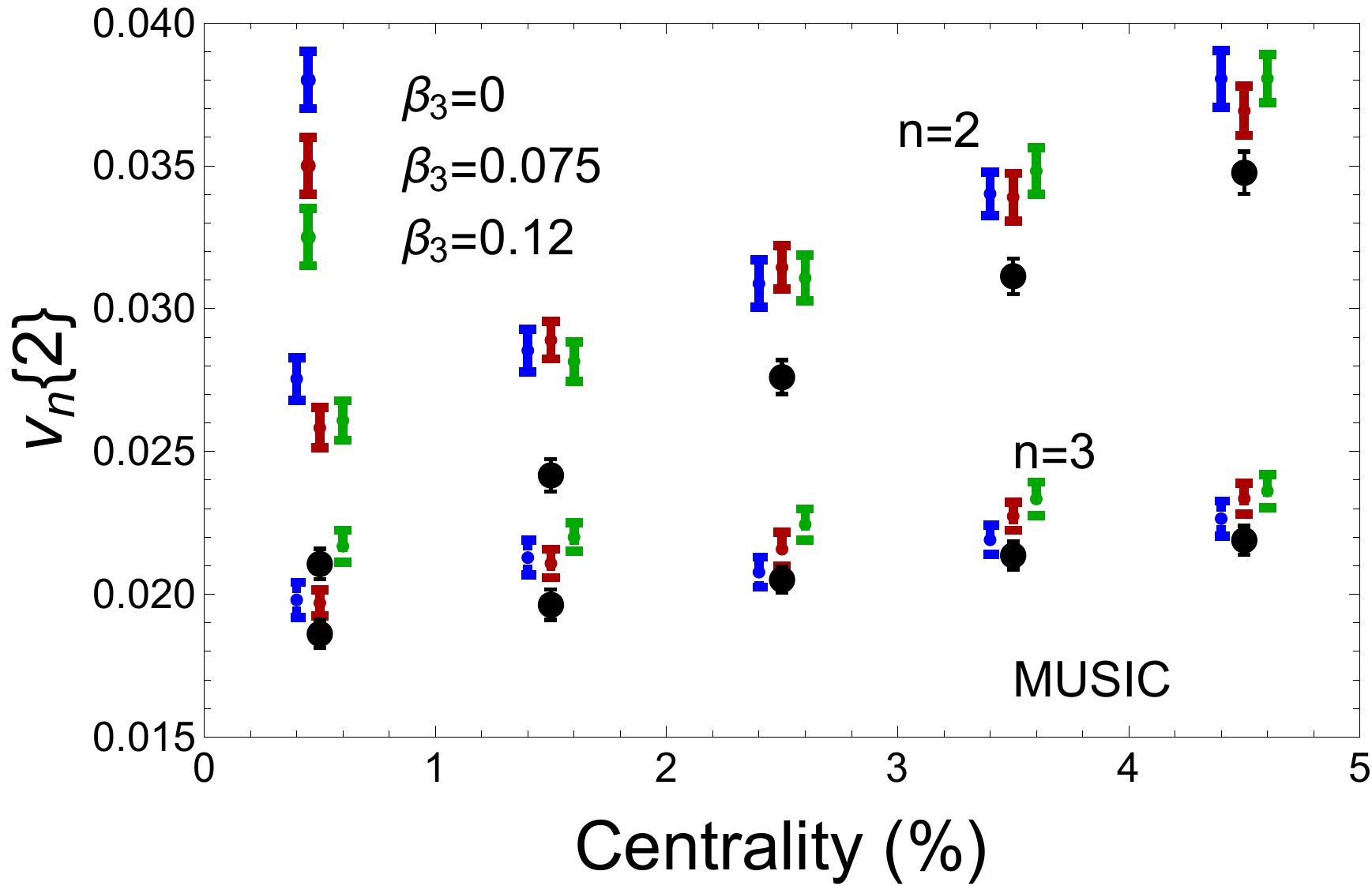} \\
  \includegraphics[width=\linewidth]{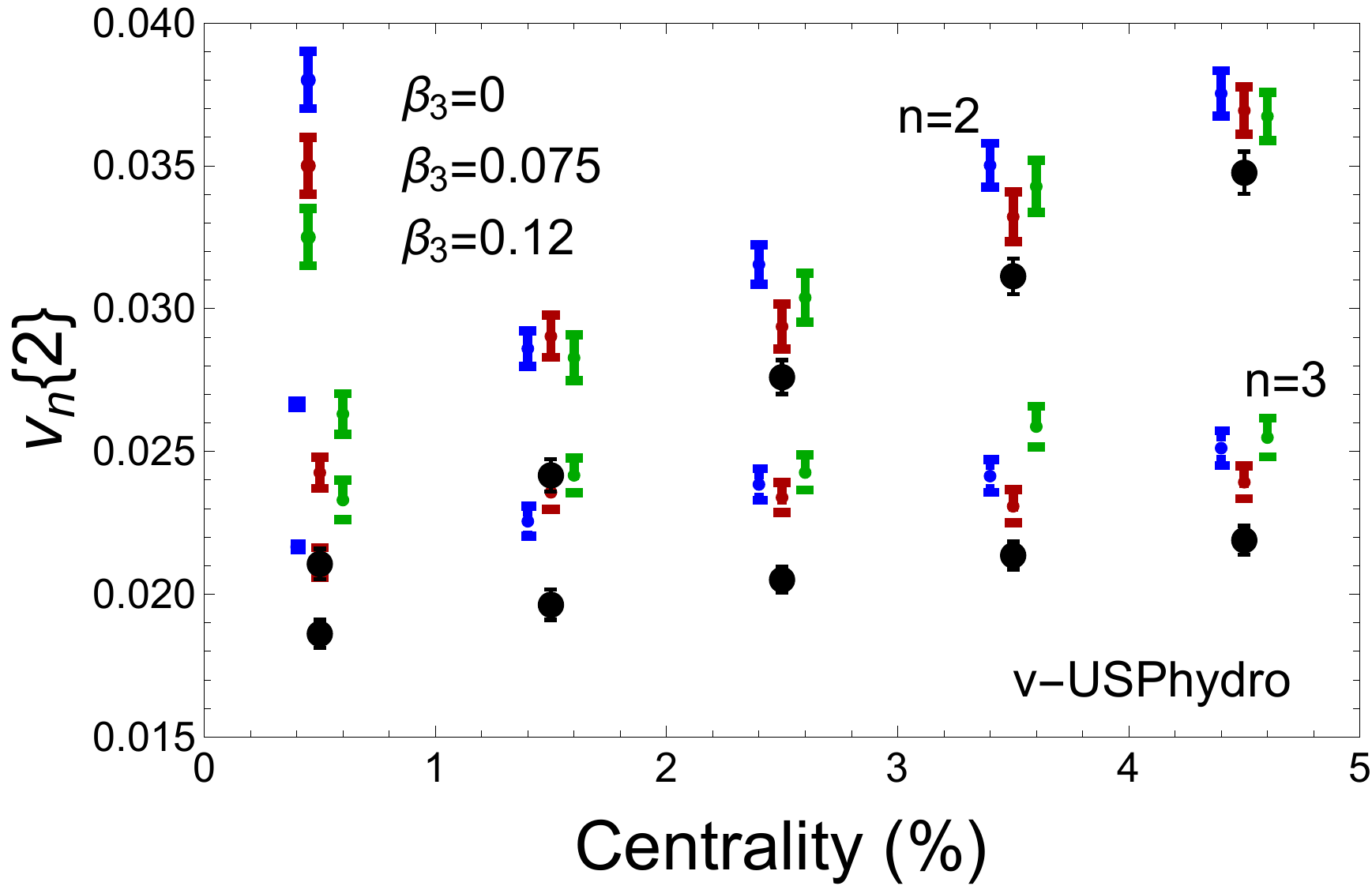} 
\caption{(Color online)  Absolute values of $v_2\{2\}$ and $v_3\{2\}$ in PbPb 5.02TeV collisions from v-USPhydro and MUSIC compared to experimental data from ALICE \cite{Acharya:2018lmh, Acharya:2019vdf} varying $\beta_3$.}
\label{fig:both_b3}
\end{figure}

Indeed, this is what's seen in Fig.\ \ref{fig:both_b3} when we consider the impact of the $\beta_3$ deformation.  
While a larger $\beta_3$ generally leads to a larger $v_3$ in most cases, there appear to be some non-monotonic fluctuations, and the $\beta_3$ deformation can non-trivially affect $v_2$ as well.

\begin{figure}[ht]
\includegraphics[width=0.48\textwidth]{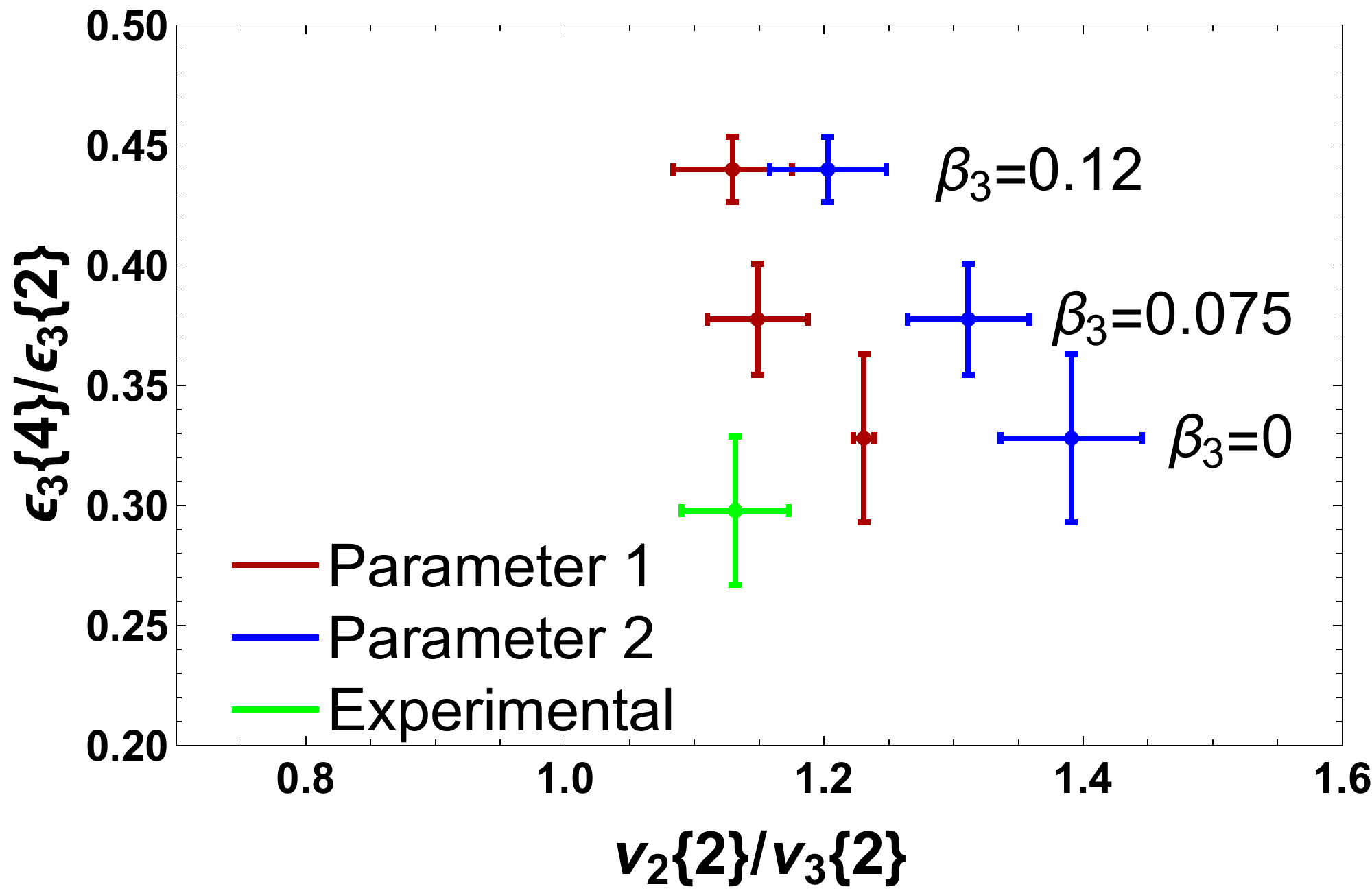}\\
\caption{Influence of the octupole deformation $\beta_3$ on multiple observables. Parameter 1 is the ratio of $\varepsilon_3\{4\} \, / \, \varepsilon_3\{2\}$ from Trento initial conditions  with $\varepsilon_2\{2\} \, / \, \varepsilon_3\{2\}$ the parameter tune from vUSPhydro. Parameter 2 is the ratio of $\varepsilon_3\{4\} \, / \, \varepsilon_3\{2\}$ from Trento initial conditions with   $\varepsilon_2\{2\} \, / \, \varepsilon_3\{2\}$ with the parameter tune from MUSIC. Clearly adding a nonzero $\beta_3$ does not improve the overall agreement with data.  Details of this plot are discussed in the text.}
\label{f:2obs}
\end{figure}

As was noted previously in Sec.~\ref{sec:ecc} and plotted in Fig.~\ref{fig:v34v32}, increasing the octupole deformation $\beta_3$ does not improve the agreement of the calculated $\varepsilon_3\{4\} \, / \, \varepsilon_3\{2\}$ with data.  Instead, the increasing disagreement with the data allows us to set an upper bound before this ratio becomes incompatible with the data.  Thus any improvement caused by $\beta_3$ in the ratio $v_2 \{2\} \, / \, v_3 \{2\}$ in Fig.~\ref{fig:v2/v3} must be weighed against the deterioration of the agreement of $v_3 \{4\} \, / \, v_3\{2\}$.  We illustrate this in Fig.~\ref{f:2obs} by plotting the two observables simultaneously on independent axes.  There are a number of caveats to this comparison which should be immediately pointed out.  The ``experimental'' point here represents a combination of the ATLAS data for $v_3 \{4\} \, / \, v_3 \{2\}$ \cite{Aad:2014vba} and the ALICE data for $v_2 \{2\} \, / \, v_3 \{2\}$ \cite{Acharya:2018lmh, Acharya:2019vdf}.  On the theory side, the points represent combinations of the initial-state $\varepsilon_3 \{4\} \, / \, \varepsilon_3 \{2\}$ (without running hydrodynamics) with the full final-state evaluation of $v_2 \{2\} \, / \, v_3 \{2\}$ for different values of the deformation $\beta_3$ and for the two hydrodynamic models.  Despite these caveats, the plot clearly illustrates the inability of an octupole deformation $\beta_3$ to resolve the $v_2$-to-$v_3$ puzzle: the improved agreement in $v_2 \{2\} \, / \, v_3 \{2\}$ is more than offset by the worsened agreement in $v_3 \{4\} \, / \, v_3 \{2\}$.

\section{Conclusions}
\label{sec:concl}

By this point, a number of possibilities to resolve the $v_2$-to-$v_3$ puzzle have been explored and exhausted.  Variations in the initial conditions (assuming spherical Pb nuclei) \cite{Shen:2015qta}, the transport coefficients \cite{Rose:2014fba}, and the equation of state \cite{Alba:2017hhe} have all been considered, yet the problem still remains.  No initial condition models assuming spherical Pb ions managed to capture the triangular flow fluctuations in ultra central PbPb collisions despite being able to reproduce elliptic flow fluctuations  \cite{Giacalone:2017uqx}.

In this paper, we considered the possibility of a nonzero octupole deformation $\beta_3$ in ${}^{208}$Pb as a means of improving the mismatch in the ratio $v_2 \{2\} \, / \, v_3 \{2\}$ in data.  Those results from full hydrodynamics (Fig.~\ref{fig:v2/v3}) and the cumulant ratio $\varepsilon_3 \{4\} \, / \, \varepsilon_3 \{2\}$ from the initial state (Fig.~\ref{fig:v34v32}) are the main results of this work.  We find that, while a finite $\beta_3$ does improve the agreement in $v_2 \{2\} \, / \, v_3 \{2\}$ with data, it simultaneously worsens the agreement in $v_3 \{4\} \, / \, v_3 \{2\}$.  Moreover, we have also performed direct comparisons to the absolute magnitudes of $v_2 \{2\}$ and $v_3 \{2\}$, rather than just their ratio (Fig.~\ref{fig:both_b3}).  We find that, in both hydrodynamic models, the biggest discrepancy underlying the $v_2$-to-$v_3$ puzzle lies with an overprediction of $v_2 \{2\}$ rather than the underprediction of $v_3 \{2\}$.  

Clearly, an octupole deformation of ${}^{208}$Pb is not a viable resolution of the $v_2$-to-$v_3$ puzzle in ultracentral collisions within these models.  While different in the details of their implementation and emphasizing different physics, both v-USPhydro and MUSIC are examples of reasonable hydrodynamic models that can describe bulk flow observables in heavy ion collisions.  The two hydrodynamic models differ in various details, but in particular the larger viscosity in MUSIC (see Fig.~\ref{f:viscosities}) leads to noticeable differences.  That larger viscosity creates additional viscous suppression of higher harmonics like $v_3$, bringing MUSIC into better quantitative agreement with $v_3 \{2\}$ than v-USPhydro, which overpredicts the data.

This difference arising from the viscosity illustrates an important correlation between the parameters like $\eta / s$ entering the hydrodynamics models, and parameters like $\beta_3$ entering the initial state.  By adjusting $\beta_3$, one can dial the ratio $v_2 \{2\} \, / \, v_3 \{2\}$ but not control the absolute magnitudes of either quantity.  Those magnitudes are affected directly by the choice of viscosity, in correlation with the prior choice of $\beta_3$.  Thus it seems reasonable that a simultaneous fit of the unknown properties of the initial state (such as the deformation parameters $\beta_2$, $\beta_3$) and the hydrodynamic transport parameters could help to resolve the $v_2$-to-$v_3$ puzzle.

For the hydrodynamic models, clearly the choice of viscosity has a significant effect on the $v_2$-to-$v_3$ puzzle, possibly indicating that a more sophisticated treatment of out-of-equilibrium effects could be important.  Adding a temperature-dependent $\eta / s$ to the v-USPhydro model could also be an important improvement.

Following our analysis, we stress that studying the $v_2$-to-$v_3$ puzzle only through the ratio $\frac{v_2\{2\}}{v_3\{2\}}$ can be highly misleading.  Independent observables, like the multiparticle correlations $\frac{v_3\{4\}}{v_3\{2\}}$, are vital to constrain the origin of these discrepancies.  Additionally, it is critical to separately examine the absolute magnitudes of $v_2\{2\}$ and $v_3\{2\}$ as the baseline for the ratio $\frac{v_2\{2\}}{v_3\{2\}}$.  For now, the $v_2$-to-$v_3$ puzzle in ultracentral Pb Pb collisions remains a challenge for hydrodynamic models.

\section*{Acknowledgements}
J.N.H., M.S., and P.C. acknowledge support from the US-DOE Nuclear Science Grant No. DE-SC0019175, the Alfred P. Sloan Foundation, and the Illinois Campus Cluster, a computing resource that is operated by the Illinois Campus Cluster Program (ICCP) in conjunction with the National Center for Supercomputing Applications (NCSA) and which is supported by funds from the University of Illinois at Urbana-Champaign.  M.L.~acknowledges support  by FAPESP projects 2016/24029-6, 2017/05685-2 and 2018/24720-6 and by project INCT-FNA Proc.~No.~464898/2014-5.


\appendix 

\begin{figure}[htb]
\includegraphics[trim=0 200 0 200,clip,width=\linewidth]{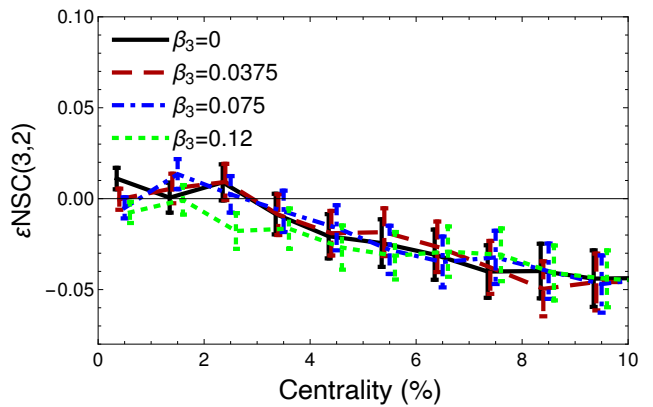} 
\caption{(Color online)  Normalized symmetric cumulants $\varepsilon NSC(3,2)$ of the initial-state eccentricities in central collisions for PbPb 5.02TeV for varying octupole deformations, $\beta_3$.}
\label{fig:eNSC}
\end{figure}

%
\section{Other Observables}
%

\subsection{Symmetric Cumulant $\varepsilon NSC(3,2)$}

As part of the $v_2$-to-$v_3$ puzzle, it is natural to ask whether the correlation between these quantities, the symmetric cumulant $NSC(3,2)$, is sensitive to a potential deformation.  In the initial state, the analogous quantity is the correlation $\varepsilon NSC(3,2)$ among the eccentricities $\varepsilon_2$ and $\varepsilon_3$, which was shown in Refs.~\cite{ALICE:2016kpq,Giacalone:2016afq, Gardim:2016nrr} to be closely related to the final-state $NSC(3,2)$. 

We plot the symmetric cumulant $NSC(3,2)$ in Fig. \ref{fig:eNSC} for a range of octupole deformations $\beta_3$.  Even with 3 million events we find that all  $\beta_3$ variations are indistinguishable within the statistical error bars.  Thus, we find that $NSC(3,2)$ is not an observable that is sensitive to deformations. 

\subsection{Linear Response Pearson Coefficients $Q_n$}

It is clear from the discussion of Fig.~\ref{fig:both_b3} that v-USPhydro and MUSIC produce different final flow harmonics. This is due to differences in their ``best fit" medium parameters --- primarily viscosity, although they also differ slightly in other respects, such as the equation of state as well.  In ultracentral collisions, both models are well described by linear response theory, with the model differences yielding different constants of proportionality.  Aside from this, the two models also may receive small contributions from nonlinear response, and these contributions can be different.  The deviations from linear response are encapsulated in the linear (Pearson) correlation coefficients $Q_n$ between initial eccentricity $\varepsilon_n$ and the final flow harmonic $v_n$.

\begin{figure}[htb]
 \includegraphics[width=\linewidth]{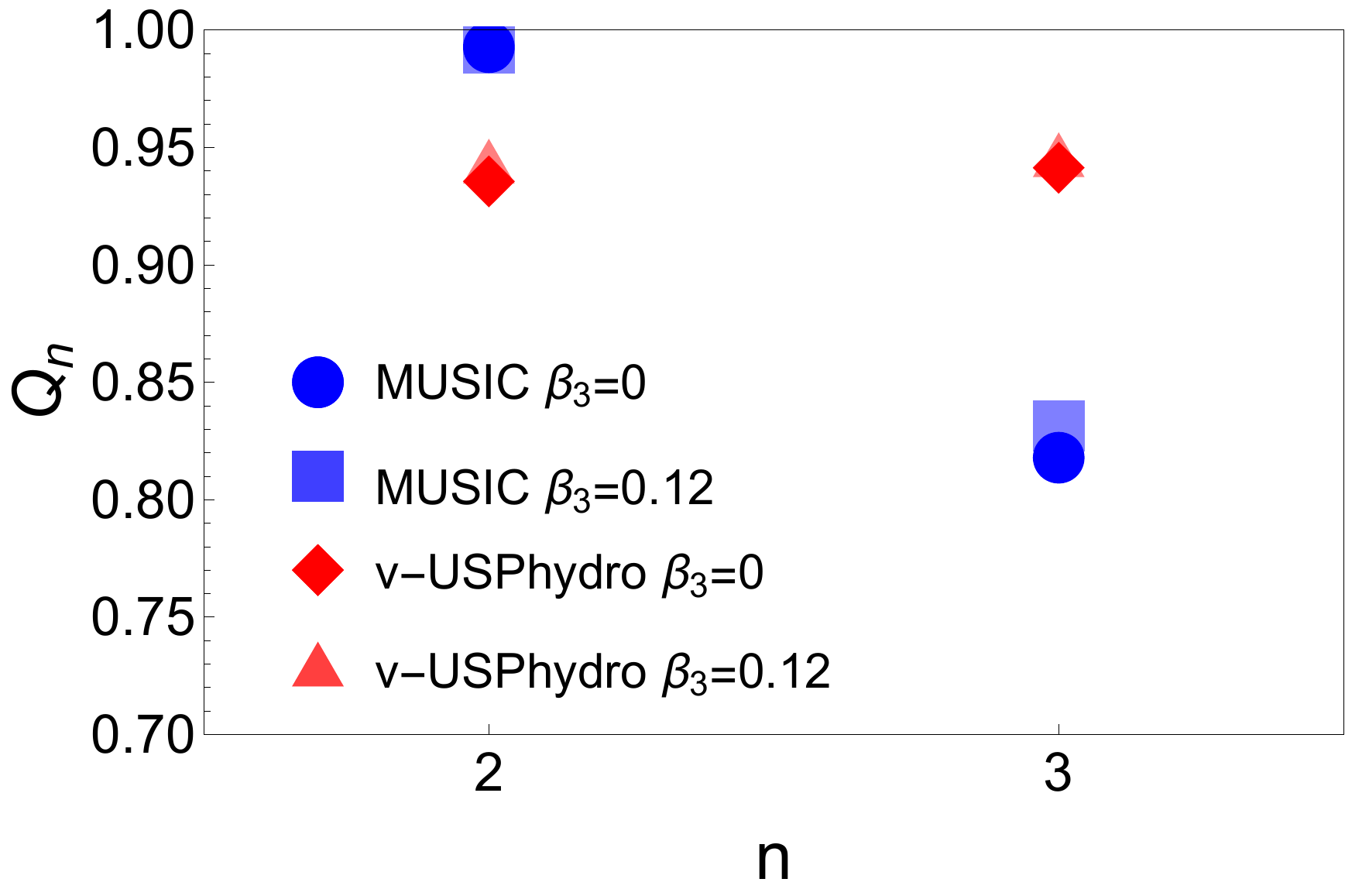} 
\caption{(Color online)  Pearson coefficients for $v_2$ and $v_3$ from MUSIC and v-USPhydro depending on the deformation.}
\label{fig:pear}
\end{figure}

We find in Fig.\ \ref{fig:pear} that the MUSIC simulation has a very strong linear relation \eqref{eq:vnepsn}, event-by-event for $v_2$, but a weaker linear correlation for $v_3$. Conversely, the v-USPhydro show a similar linear correlation for both harmonics.  Without a dedicated study, it is difficult to further disentangle how the various features of the models result in the overall mapping from initial state eccentricities to final state flow harmonics.

\section*{References}
\bibliography{allv3}{}

\begin{thebibliography}{80}%
\makeatletter
\providecommand \@ifxundefined [1]{%
 \@ifx{#1\undefined}
}%
\providecommand \@ifnum [1]{%
 \ifnum #1\expandafter \@firstoftwo
 \else \expandafter \@secondoftwo
 \fi
}%
\providecommand \@ifx [1]{%
 \ifx #1\expandafter \@firstoftwo
 \else \expandafter \@secondoftwo
 \fi
}%
\providecommand \natexlab [1]{#1}%
\providecommand \enquote  [1]{``#1''}%
\providecommand \bibnamefont  [1]{#1}%
\providecommand \bibfnamefont [1]{#1}%
\providecommand \citenamefont [1]{#1}%
\providecommand \href@noop [0]{\@secondoftwo}%
\providecommand \href [0]{\begingroup \@sanitize@url \@href}%
\providecommand \@href[1]{\@@startlink{#1}\@@href}%
\providecommand \@@href[1]{\endgroup#1\@@endlink}%
\providecommand \@sanitize@url [0]{\catcode `\\12\catcode `\$12\catcode
  `\&12\catcode `\#12\catcode `\^12\catcode `\_12\catcode `\%12\relax}%
\providecommand \@@startlink[1]{}%
\providecommand \@@endlink[0]{}%
\providecommand \url  [0]{\begingroup\@sanitize@url \@url }%
\providecommand \@url [1]{\endgroup\@href {#1}{\urlprefix }}%
\providecommand \urlprefix  [0]{URL }%
\providecommand \Eprint [0]{\href }%
\providecommand \doibase [0]{http://dx.doi.org/}%
\providecommand \selectlanguage [0]{\@gobble}%
\providecommand \bibinfo  [0]{\@secondoftwo}%
\providecommand \bibfield  [0]{\@secondoftwo}%
\providecommand \translation [1]{[#1]}%
\providecommand \BibitemOpen [0]{}%
\providecommand \bibitemStop [0]{}%
\providecommand \bibitemNoStop [0]{.\EOS\space}%
\providecommand \EOS [0]{\spacefactor3000\relax}%
\providecommand \BibitemShut  [1]{\csname bibitem#1\endcsname}%
\let\auto@bib@innerbib\@empty
\bibitem [{\citenamefont {Noronha-Hostler}\ \emph
  {et~al.}(2016{\natexlab{a}})\citenamefont {Noronha-Hostler}, \citenamefont
  {Luzum},\ and\ \citenamefont {Ollitrault}}]{Noronha-Hostler:2015uye}%
  \BibitemOpen
  \bibfield  {author} {\bibinfo {author} {\bibfnamefont {J.}~\bibnamefont
  {Noronha-Hostler}}, \bibinfo {author} {\bibfnamefont {M.}~\bibnamefont
  {Luzum}}, \ and\ \bibinfo {author} {\bibfnamefont {J.-Y.}\ \bibnamefont
  {Ollitrault}},\ }\href {\doibase 10.1103/PhysRevC.93.034912} {\bibfield
  {journal} {\bibinfo  {journal} {Phys. Rev.}\ }\textbf {\bibinfo {volume}
  {C93}},\ \bibinfo {pages} {034912} (\bibinfo {year} {2016}{\natexlab{a}})},\
  \Eprint {http://arxiv.org/abs/1511.06289} {arXiv:1511.06289 [nucl-th]}
  \BibitemShut {NoStop}%
\bibitem [{\citenamefont {Niemi}\ \emph
  {et~al.}(2016{\natexlab{a}})\citenamefont {Niemi}, \citenamefont {Eskola},
  \citenamefont {Paatelainen},\ and\ \citenamefont {Tuominen}}]{Niemi:2015voa}%
  \BibitemOpen
  \bibfield  {author} {\bibinfo {author} {\bibfnamefont {H.}~\bibnamefont
  {Niemi}}, \bibinfo {author} {\bibfnamefont {K.~J.}\ \bibnamefont {Eskola}},
  \bibinfo {author} {\bibfnamefont {R.}~\bibnamefont {Paatelainen}}, \ and\
  \bibinfo {author} {\bibfnamefont {K.}~\bibnamefont {Tuominen}},\ }\href
  {\doibase 10.1103/PhysRevC.93.014912} {\bibfield  {journal} {\bibinfo
  {journal} {Phys. Rev.}\ }\textbf {\bibinfo {volume} {C93}},\ \bibinfo {pages}
  {014912} (\bibinfo {year} {2016}{\natexlab{a}})},\ \Eprint
  {http://arxiv.org/abs/1511.04296} {arXiv:1511.04296 [hep-ph]} \BibitemShut
  {NoStop}%
\bibitem [{\citenamefont {Adam}\ \emph
  {et~al.}(2016{\natexlab{a}})\citenamefont {Adam} \emph
  {et~al.}}]{Adam:2015ptt}%
  \BibitemOpen
  \bibfield  {author} {\bibinfo {author} {\bibfnamefont {J.}~\bibnamefont
  {Adam}} \emph {et~al.} (\bibinfo {collaboration} {ALICE}),\ }\href {\doibase
  10.1103/PhysRevLett.116.222302} {\bibfield  {journal} {\bibinfo  {journal}
  {Phys. Rev. Lett.}\ }\textbf {\bibinfo {volume} {116}},\ \bibinfo {pages}
  {222302} (\bibinfo {year} {2016}{\natexlab{a}})},\ \Eprint
  {http://arxiv.org/abs/1512.06104} {arXiv:1512.06104 [nucl-ex]} \BibitemShut
  {NoStop}%
\bibitem [{\citenamefont {Song}\ \emph {et~al.}(2011)\citenamefont {Song},
  \citenamefont {Bass}, \citenamefont {Heinz}, \citenamefont {Hirano},\ and\
  \citenamefont {Shen}}]{Song:2010mg}%
  \BibitemOpen
  \bibfield  {author} {\bibinfo {author} {\bibfnamefont {H.}~\bibnamefont
  {Song}}, \bibinfo {author} {\bibfnamefont {S.~A.}\ \bibnamefont {Bass}},
  \bibinfo {author} {\bibfnamefont {U.}~\bibnamefont {Heinz}}, \bibinfo
  {author} {\bibfnamefont {T.}~\bibnamefont {Hirano}}, \ and\ \bibinfo {author}
  {\bibfnamefont {C.}~\bibnamefont {Shen}},\ }\href {\doibase
  10.1103/PhysRevLett.106.192301, 10.1103/PhysRevLett.109.139904} {\bibfield
  {journal} {\bibinfo  {journal} {Phys. Rev. Lett.}\ }\textbf {\bibinfo
  {volume} {106}},\ \bibinfo {pages} {192301} (\bibinfo {year} {2011})},\
  \bibinfo {note} {[Erratum: Phys. Rev. Lett.109,139904(2012)]},\ \Eprint
  {http://arxiv.org/abs/1011.2783} {arXiv:1011.2783 [nucl-th]} \BibitemShut
  {NoStop}%
\bibitem [{\citenamefont {Bozek}\ and\ \citenamefont
  {Wyskiel-Piekarska}(2012)}]{Bozek:2012qs}%
  \BibitemOpen
  \bibfield  {author} {\bibinfo {author} {\bibfnamefont {P.}~\bibnamefont
  {Bozek}}\ and\ \bibinfo {author} {\bibfnamefont {I.}~\bibnamefont
  {Wyskiel-Piekarska}},\ }\href {\doibase 10.1103/PhysRevC.85.064915}
  {\bibfield  {journal} {\bibinfo  {journal} {Phys. Rev.}\ }\textbf {\bibinfo
  {volume} {C85}},\ \bibinfo {pages} {064915} (\bibinfo {year} {2012})},\
  \Eprint {http://arxiv.org/abs/1203.6513} {arXiv:1203.6513 [nucl-th]}
  \BibitemShut {NoStop}%
\bibitem [{\citenamefont {Gardim}\ \emph {et~al.}(2012)\citenamefont {Gardim},
  \citenamefont {Grassi}, \citenamefont {Luzum},\ and\ \citenamefont
  {Ollitrault}}]{Gardim:2012yp}%
  \BibitemOpen
  \bibfield  {author} {\bibinfo {author} {\bibfnamefont {F.~G.}\ \bibnamefont
  {Gardim}}, \bibinfo {author} {\bibfnamefont {F.}~\bibnamefont {Grassi}},
  \bibinfo {author} {\bibfnamefont {M.}~\bibnamefont {Luzum}}, \ and\ \bibinfo
  {author} {\bibfnamefont {J.-Y.}\ \bibnamefont {Ollitrault}},\ }\href
  {\doibase 10.1103/PhysRevLett.109.202302} {\bibfield  {journal} {\bibinfo
  {journal} {Phys. Rev. Lett.}\ }\textbf {\bibinfo {volume} {109}},\ \bibinfo
  {pages} {202302} (\bibinfo {year} {2012})},\ \Eprint
  {http://arxiv.org/abs/1203.2882} {arXiv:1203.2882 [nucl-th]} \BibitemShut
  {NoStop}%
\bibitem [{\citenamefont {Bozek}\ and\ \citenamefont
  {Broniowski}(2013)}]{Bozek:2013uha}%
  \BibitemOpen
  \bibfield  {author} {\bibinfo {author} {\bibfnamefont {P.}~\bibnamefont
  {Bozek}}\ and\ \bibinfo {author} {\bibfnamefont {W.}~\bibnamefont
  {Broniowski}},\ }\href {\doibase 10.1103/PhysRevC.88.014903} {\bibfield
  {journal} {\bibinfo  {journal} {Phys. Rev.}\ }\textbf {\bibinfo {volume}
  {C88}},\ \bibinfo {pages} {014903} (\bibinfo {year} {2013})},\ \Eprint
  {http://arxiv.org/abs/1304.3044} {arXiv:1304.3044 [nucl-th]} \BibitemShut
  {NoStop}%
\bibitem [{\citenamefont {Niemi}\ \emph
  {et~al.}(2016{\natexlab{b}})\citenamefont {Niemi}, \citenamefont {Eskola},\
  and\ \citenamefont {Paatelainen}}]{Niemi:2015qia}%
  \BibitemOpen
  \bibfield  {author} {\bibinfo {author} {\bibfnamefont {H.}~\bibnamefont
  {Niemi}}, \bibinfo {author} {\bibfnamefont {K.~J.}\ \bibnamefont {Eskola}}, \
  and\ \bibinfo {author} {\bibfnamefont {R.}~\bibnamefont {Paatelainen}},\
  }\href {\doibase 10.1103/PhysRevC.93.024907} {\bibfield  {journal} {\bibinfo
  {journal} {Phys. Rev.}\ }\textbf {\bibinfo {volume} {C93}},\ \bibinfo {pages}
  {024907} (\bibinfo {year} {2016}{\natexlab{b}})},\ \Eprint
  {http://arxiv.org/abs/1505.02677} {arXiv:1505.02677 [hep-ph]} \BibitemShut
  {NoStop}%
\bibitem [{\citenamefont {Ryu}\ \emph {et~al.}(2015)\citenamefont {Ryu},
  \citenamefont {Paquet}, \citenamefont {Shen}, \citenamefont {Denicol},
  \citenamefont {Schenke}, \citenamefont {Jeon},\ and\ \citenamefont
  {Gale}}]{Ryu:2015vwa}%
  \BibitemOpen
  \bibfield  {author} {\bibinfo {author} {\bibfnamefont {S.}~\bibnamefont
  {Ryu}}, \bibinfo {author} {\bibfnamefont {J.~F.}\ \bibnamefont {Paquet}},
  \bibinfo {author} {\bibfnamefont {C.}~\bibnamefont {Shen}}, \bibinfo {author}
  {\bibfnamefont {G.~S.}\ \bibnamefont {Denicol}}, \bibinfo {author}
  {\bibfnamefont {B.}~\bibnamefont {Schenke}}, \bibinfo {author} {\bibfnamefont
  {S.}~\bibnamefont {Jeon}}, \ and\ \bibinfo {author} {\bibfnamefont
  {C.}~\bibnamefont {Gale}},\ }\href {\doibase 10.1103/PhysRevLett.115.132301}
  {\bibfield  {journal} {\bibinfo  {journal} {Phys. Rev. Lett.}\ }\textbf
  {\bibinfo {volume} {115}},\ \bibinfo {pages} {132301} (\bibinfo {year}
  {2015})},\ \Eprint {http://arxiv.org/abs/1502.01675} {arXiv:1502.01675
  [nucl-th]} \BibitemShut {NoStop}%
\bibitem [{\citenamefont {McDonald}\ \emph {et~al.}(2017)\citenamefont
  {McDonald}, \citenamefont {Shen}, \citenamefont {Fillion-Gourdeau},
  \citenamefont {Jeon},\ and\ \citenamefont {Gale}}]{McDonald:2016vlt}%
  \BibitemOpen
  \bibfield  {author} {\bibinfo {author} {\bibfnamefont {S.}~\bibnamefont
  {McDonald}}, \bibinfo {author} {\bibfnamefont {C.}~\bibnamefont {Shen}},
  \bibinfo {author} {\bibfnamefont {F.}~\bibnamefont {Fillion-Gourdeau}},
  \bibinfo {author} {\bibfnamefont {S.}~\bibnamefont {Jeon}}, \ and\ \bibinfo
  {author} {\bibfnamefont {C.}~\bibnamefont {Gale}},\ }\href {\doibase
  10.1103/PhysRevC.95.064913} {\bibfield  {journal} {\bibinfo  {journal} {Phys.
  Rev.}\ }\textbf {\bibinfo {volume} {C95}},\ \bibinfo {pages} {064913}
  (\bibinfo {year} {2017})},\ \Eprint {http://arxiv.org/abs/1609.02958}
  {arXiv:1609.02958 [hep-ph]} \BibitemShut {NoStop}%
\bibitem [{\citenamefont {Bernhard}\ \emph {et~al.}(2016)\citenamefont
  {Bernhard}, \citenamefont {Moreland}, \citenamefont {Bass}, \citenamefont
  {Liu},\ and\ \citenamefont {Heinz}}]{Bernhard:2016tnd}%
  \BibitemOpen
  \bibfield  {author} {\bibinfo {author} {\bibfnamefont {J.~E.}\ \bibnamefont
  {Bernhard}}, \bibinfo {author} {\bibfnamefont {J.~S.}\ \bibnamefont
  {Moreland}}, \bibinfo {author} {\bibfnamefont {S.~A.}\ \bibnamefont {Bass}},
  \bibinfo {author} {\bibfnamefont {J.}~\bibnamefont {Liu}}, \ and\ \bibinfo
  {author} {\bibfnamefont {U.}~\bibnamefont {Heinz}},\ }\href {\doibase
  10.1103/PhysRevC.94.024907} {\bibfield  {journal} {\bibinfo  {journal} {Phys.
  Rev.}\ }\textbf {\bibinfo {volume} {C94}},\ \bibinfo {pages} {024907}
  (\bibinfo {year} {2016})},\ \Eprint {http://arxiv.org/abs/1605.03954}
  {arXiv:1605.03954 [nucl-th]} \BibitemShut {NoStop}%
\bibitem [{\citenamefont {Gardim}\ \emph {et~al.}(2017)\citenamefont {Gardim},
  \citenamefont {Grassi}, \citenamefont {Luzum},\ and\ \citenamefont
  {Noronha-Hostler}}]{Gardim:2016nrr}%
  \BibitemOpen
  \bibfield  {author} {\bibinfo {author} {\bibfnamefont {F.~G.}\ \bibnamefont
  {Gardim}}, \bibinfo {author} {\bibfnamefont {F.}~\bibnamefont {Grassi}},
  \bibinfo {author} {\bibfnamefont {M.}~\bibnamefont {Luzum}}, \ and\ \bibinfo
  {author} {\bibfnamefont {J.}~\bibnamefont {Noronha-Hostler}},\ }\href
  {\doibase 10.1103/PhysRevC.95.034901} {\bibfield  {journal} {\bibinfo
  {journal} {Phys. Rev.}\ }\textbf {\bibinfo {volume} {C95}},\ \bibinfo {pages}
  {034901} (\bibinfo {year} {2017})},\ \Eprint
  {http://arxiv.org/abs/1608.02982} {arXiv:1608.02982 [nucl-th]} \BibitemShut
  {NoStop}%
\bibitem [{\citenamefont {Alba}\ \emph {et~al.}(2018)\citenamefont {Alba},
  \citenamefont {Mantovani~Sarti}, \citenamefont {Noronha}, \citenamefont
  {Noronha-Hostler}, \citenamefont {Parotto}, \citenamefont
  {Portillo~Vazquez},\ and\ \citenamefont {Ratti}}]{Alba:2017hhe}%
  \BibitemOpen
  \bibfield  {author} {\bibinfo {author} {\bibfnamefont {P.}~\bibnamefont
  {Alba}}, \bibinfo {author} {\bibfnamefont {V.}~\bibnamefont
  {Mantovani~Sarti}}, \bibinfo {author} {\bibfnamefont {J.}~\bibnamefont
  {Noronha}}, \bibinfo {author} {\bibfnamefont {J.}~\bibnamefont
  {Noronha-Hostler}}, \bibinfo {author} {\bibfnamefont {P.}~\bibnamefont
  {Parotto}}, \bibinfo {author} {\bibfnamefont {I.}~\bibnamefont
  {Portillo~Vazquez}}, \ and\ \bibinfo {author} {\bibfnamefont
  {C.}~\bibnamefont {Ratti}},\ }\href {\doibase 10.1103/PhysRevC.98.034909}
  {\bibfield  {journal} {\bibinfo  {journal} {Phys. Rev.}\ }\textbf {\bibinfo
  {volume} {C98}},\ \bibinfo {pages} {034909} (\bibinfo {year} {2018})},\
  \Eprint {http://arxiv.org/abs/1711.05207} {arXiv:1711.05207 [nucl-th]}
  \BibitemShut {NoStop}%
\bibitem [{\citenamefont {Giacalone}\ \emph {et~al.}(2018)\citenamefont
  {Giacalone}, \citenamefont {Noronha-Hostler}, \citenamefont {Luzum},\ and\
  \citenamefont {Ollitrault}}]{Giacalone:2017dud}%
  \BibitemOpen
  \bibfield  {author} {\bibinfo {author} {\bibfnamefont {G.}~\bibnamefont
  {Giacalone}}, \bibinfo {author} {\bibfnamefont {J.}~\bibnamefont
  {Noronha-Hostler}}, \bibinfo {author} {\bibfnamefont {M.}~\bibnamefont
  {Luzum}}, \ and\ \bibinfo {author} {\bibfnamefont {J.-Y.}\ \bibnamefont
  {Ollitrault}},\ }\href {\doibase 10.1103/PhysRevC.97.034904} {\bibfield
  {journal} {\bibinfo  {journal} {Phys. Rev.}\ }\textbf {\bibinfo {volume}
  {C97}},\ \bibinfo {pages} {034904} (\bibinfo {year} {2018})},\ \Eprint
  {http://arxiv.org/abs/1711.08499} {arXiv:1711.08499 [nucl-th]} \BibitemShut
  {NoStop}%
\bibitem [{\citenamefont {Eskola}\ \emph {et~al.}(2018)\citenamefont {Eskola},
  \citenamefont {Niemi}, \citenamefont {Paatelainen},\ and\ \citenamefont
  {Tuominen}}]{Eskola:2017bup}%
  \BibitemOpen
  \bibfield  {author} {\bibinfo {author} {\bibfnamefont {K.~J.}\ \bibnamefont
  {Eskola}}, \bibinfo {author} {\bibfnamefont {H.}~\bibnamefont {Niemi}},
  \bibinfo {author} {\bibfnamefont {R.}~\bibnamefont {Paatelainen}}, \ and\
  \bibinfo {author} {\bibfnamefont {K.}~\bibnamefont {Tuominen}},\ }\href
  {\doibase 10.1103/PhysRevC.97.034911} {\bibfield  {journal} {\bibinfo
  {journal} {Phys. Rev.}\ }\textbf {\bibinfo {volume} {C97}},\ \bibinfo {pages}
  {034911} (\bibinfo {year} {2018})},\ \Eprint
  {http://arxiv.org/abs/1711.09803} {arXiv:1711.09803 [hep-ph]} \BibitemShut
  {NoStop}%
\bibitem [{\citenamefont {Weller}\ and\ \citenamefont
  {Romatschke}(2017)}]{Weller:2017tsr}%
  \BibitemOpen
  \bibfield  {author} {\bibinfo {author} {\bibfnamefont {R.~D.}\ \bibnamefont
  {Weller}}\ and\ \bibinfo {author} {\bibfnamefont {P.}~\bibnamefont
  {Romatschke}},\ }\href {\doibase 10.1016/j.physletb.2017.09.077} {\bibfield
  {journal} {\bibinfo  {journal} {Phys. Lett.}\ }\textbf {\bibinfo {volume}
  {B774}},\ \bibinfo {pages} {351} (\bibinfo {year} {2017})},\ \Eprint
  {http://arxiv.org/abs/1701.07145} {arXiv:1701.07145 [nucl-th]} \BibitemShut
  {NoStop}%
\bibitem [{\citenamefont {Schenke}\ \emph {et~al.}(2019)\citenamefont
  {Schenke}, \citenamefont {Shen},\ and\ \citenamefont
  {Tribedy}}]{Schenke:2019ruo}%
  \BibitemOpen
  \bibfield  {author} {\bibinfo {author} {\bibfnamefont {B.}~\bibnamefont
  {Schenke}}, \bibinfo {author} {\bibfnamefont {C.}~\bibnamefont {Shen}}, \
  and\ \bibinfo {author} {\bibfnamefont {P.}~\bibnamefont {Tribedy}},\ }\href
  {\doibase 10.1103/PhysRevC.99.044908} {\bibfield  {journal} {\bibinfo
  {journal} {Phys. Rev.}\ }\textbf {\bibinfo {volume} {C99}},\ \bibinfo {pages}
  {044908} (\bibinfo {year} {2019})},\ \Eprint
  {http://arxiv.org/abs/1901.04378} {arXiv:1901.04378 [nucl-th]} \BibitemShut
  {NoStop}%
\bibitem [{\citenamefont {Gardim}\ \emph {et~al.}(2018)\citenamefont {Gardim},
  \citenamefont {Grassi}, \citenamefont {Ishida}, \citenamefont {Luzum},
  \citenamefont {Magalhães},\ and\ \citenamefont
  {Noronha-Hostler}}]{Gardim:2017ruc}%
  \BibitemOpen
  \bibfield  {author} {\bibinfo {author} {\bibfnamefont {F.~G.}\ \bibnamefont
  {Gardim}}, \bibinfo {author} {\bibfnamefont {F.}~\bibnamefont {Grassi}},
  \bibinfo {author} {\bibfnamefont {P.}~\bibnamefont {Ishida}}, \bibinfo
  {author} {\bibfnamefont {M.}~\bibnamefont {Luzum}}, \bibinfo {author}
  {\bibfnamefont {P.~S.}\ \bibnamefont {Magalhães}}, \ and\ \bibinfo {author}
  {\bibfnamefont {J.}~\bibnamefont {Noronha-Hostler}},\ }\href {\doibase
  10.1103/PhysRevC.97.064919} {\bibfield  {journal} {\bibinfo  {journal} {Phys.
  Rev.}\ }\textbf {\bibinfo {volume} {C97}},\ \bibinfo {pages} {064919}
  (\bibinfo {year} {2018})},\ \Eprint {http://arxiv.org/abs/1712.03912}
  {arXiv:1712.03912 [nucl-th]} \BibitemShut {NoStop}%
\bibitem [{\citenamefont {Noronha-Hostler}\ \emph
  {et~al.}(2016{\natexlab{b}})\citenamefont {Noronha-Hostler}, \citenamefont
  {Noronha},\ and\ \citenamefont {Gyulassy}}]{Noronha-Hostler:2015coa}%
  \BibitemOpen
  \bibfield  {author} {\bibinfo {author} {\bibfnamefont {J.}~\bibnamefont
  {Noronha-Hostler}}, \bibinfo {author} {\bibfnamefont {J.}~\bibnamefont
  {Noronha}}, \ and\ \bibinfo {author} {\bibfnamefont {M.}~\bibnamefont
  {Gyulassy}},\ }\href {\doibase 10.1103/PhysRevC.93.024909} {\bibfield
  {journal} {\bibinfo  {journal} {Phys. Rev.}\ }\textbf {\bibinfo {volume}
  {C93}},\ \bibinfo {pages} {024909} (\bibinfo {year} {2016}{\natexlab{b}})},\
  \Eprint {http://arxiv.org/abs/1508.02455} {arXiv:1508.02455 [nucl-th]}
  \BibitemShut {NoStop}%
\bibitem [{\citenamefont {Nagle}\ and\ \citenamefont
  {Zajc}(2019)}]{Nagle:2018ybc}%
  \BibitemOpen
  \bibfield  {author} {\bibinfo {author} {\bibfnamefont {J.~L.}\ \bibnamefont
  {Nagle}}\ and\ \bibinfo {author} {\bibfnamefont {W.~A.}\ \bibnamefont
  {Zajc}},\ }\href {\doibase 10.1103/PhysRevC.99.054908} {\bibfield  {journal}
  {\bibinfo  {journal} {Phys. Rev.}\ }\textbf {\bibinfo {volume} {C99}},\
  \bibinfo {pages} {054908} (\bibinfo {year} {2019})},\ \Eprint
  {http://arxiv.org/abs/1808.01276} {arXiv:1808.01276 [nucl-th]} \BibitemShut
  {NoStop}%
\bibitem [{\citenamefont {Hippert}\ \emph {et~al.}(2020)\citenamefont
  {Hippert}, \citenamefont {Barbon}, \citenamefont {Dobrigkeit~Chinellato},
  \citenamefont {Luzum}, \citenamefont {Noronha}, \citenamefont {Nunes~da
  Silva}, \citenamefont {Serenone},\ and\ \citenamefont
  {Takahashi}}]{Hippert:2020kde}%
  \BibitemOpen
  \bibfield  {author} {\bibinfo {author} {\bibfnamefont {M.}~\bibnamefont
  {Hippert}}, \bibinfo {author} {\bibfnamefont {J.~G.~P.}\ \bibnamefont
  {Barbon}}, \bibinfo {author} {\bibfnamefont {D.}~\bibnamefont
  {Dobrigkeit~Chinellato}}, \bibinfo {author} {\bibfnamefont {M.}~\bibnamefont
  {Luzum}}, \bibinfo {author} {\bibfnamefont {J.}~\bibnamefont {Noronha}},
  \bibinfo {author} {\bibfnamefont {T.}~\bibnamefont {Nunes~da Silva}},
  \bibinfo {author} {\bibfnamefont {W.~M.}\ \bibnamefont {Serenone}}, \ and\
  \bibinfo {author} {\bibfnamefont {J.}~\bibnamefont {Takahashi}},\ }\href@noop
  {} {\  (\bibinfo {year} {2020})},\ \Eprint {http://arxiv.org/abs/2006.13358}
  {arXiv:2006.13358 [nucl-th]} \BibitemShut {NoStop}%
\bibitem [{\citenamefont {Nunes~da Silva}\ \emph {et~al.}(2020)\citenamefont
  {Nunes~da Silva}, \citenamefont {Chinellato}, \citenamefont {Hippert},
  \citenamefont {Serenone}, \citenamefont {Takahashi}, \citenamefont {Denicol},
  \citenamefont {Luzum},\ and\ \citenamefont {Noronha}}]{NunesdaSilva:2020bfs}%
  \BibitemOpen
  \bibfield  {author} {\bibinfo {author} {\bibfnamefont {T.}~\bibnamefont
  {Nunes~da Silva}}, \bibinfo {author} {\bibfnamefont {D.}~\bibnamefont
  {Chinellato}}, \bibinfo {author} {\bibfnamefont {M.}~\bibnamefont {Hippert}},
  \bibinfo {author} {\bibfnamefont {W.}~\bibnamefont {Serenone}}, \bibinfo
  {author} {\bibfnamefont {J.}~\bibnamefont {Takahashi}}, \bibinfo {author}
  {\bibfnamefont {G.~S.}\ \bibnamefont {Denicol}}, \bibinfo {author}
  {\bibfnamefont {M.}~\bibnamefont {Luzum}}, \ and\ \bibinfo {author}
  {\bibfnamefont {J.}~\bibnamefont {Noronha}},\ }\href@noop {} {\  (\bibinfo
  {year} {2020})},\ \Eprint {http://arxiv.org/abs/2006.02324} {arXiv:2006.02324
  [nucl-th]} \BibitemShut {NoStop}%
\bibitem [{\citenamefont {Acharya}\ \emph
  {et~al.}(2019{\natexlab{a}})\citenamefont {Acharya} \emph
  {et~al.}}]{Acharya:2019vdf}%
  \BibitemOpen
  \bibfield  {author} {\bibinfo {author} {\bibfnamefont {S.}~\bibnamefont
  {Acharya}} \emph {et~al.} (\bibinfo {collaboration} {ALICE}),\ }\href
  {\doibase 10.1103/PhysRevLett.123.142301} {\bibfield  {journal} {\bibinfo
  {journal} {Phys. Rev. Lett.}\ }\textbf {\bibinfo {volume} {123}},\ \bibinfo
  {pages} {142301} (\bibinfo {year} {2019}{\natexlab{a}})},\ \Eprint
  {http://arxiv.org/abs/1903.01790} {arXiv:1903.01790 [nucl-ex]} \BibitemShut
  {NoStop}%
\bibitem [{\citenamefont {Acharya}\ \emph
  {et~al.}(2018{\natexlab{a}})\citenamefont {Acharya} \emph
  {et~al.}}]{Acharya:2018ihu}%
  \BibitemOpen
  \bibfield  {author} {\bibinfo {author} {\bibfnamefont {S.}~\bibnamefont
  {Acharya}} \emph {et~al.} (\bibinfo {collaboration} {ALICE}),\ }\href
  {\doibase 10.1016/j.physletb.2018.06.059} {\bibfield  {journal} {\bibinfo
  {journal} {Phys. Lett.}\ }\textbf {\bibinfo {volume} {B784}},\ \bibinfo
  {pages} {82} (\bibinfo {year} {2018}{\natexlab{a}})},\ \Eprint
  {http://arxiv.org/abs/1805.01832} {arXiv:1805.01832 [nucl-ex]} \BibitemShut
  {NoStop}%
\bibitem [{\citenamefont {Acharya}\ \emph
  {et~al.}(2019{\natexlab{b}})\citenamefont {Acharya} \emph
  {et~al.}}]{Acharya:2018eaq}%
  \BibitemOpen
  \bibfield  {author} {\bibinfo {author} {\bibfnamefont {S.}~\bibnamefont
  {Acharya}} \emph {et~al.} (\bibinfo {collaboration} {ALICE}),\ }\href
  {\doibase 10.1016/j.physletb.2018.10.052} {\bibfield  {journal} {\bibinfo
  {journal} {Phys. Lett.}\ }\textbf {\bibinfo {volume} {B788}},\ \bibinfo
  {pages} {166} (\bibinfo {year} {2019}{\natexlab{b}})},\ \Eprint
  {http://arxiv.org/abs/1805.04399} {arXiv:1805.04399 [nucl-ex]} \BibitemShut
  {NoStop}%
\bibitem [{\citenamefont {Collaboration}(2018)}]{CMS:2018jmx}%
  \BibitemOpen
  \bibfield  {author} {\bibinfo {author} {\bibfnamefont {C.}~\bibnamefont
  {Collaboration}} (\bibinfo {collaboration} {CMS}),\ }\href@noop {} {\
  (\bibinfo {year} {2018})}\BibitemShut {NoStop}%
\bibitem [{\citenamefont {collaboration}(2018)}]{ATLAS:2018iom}%
  \BibitemOpen
  \bibfield  {author} {\bibinfo {author} {\bibfnamefont {T.~A.}\ \bibnamefont
  {collaboration}} (\bibinfo {collaboration} {ATLAS}),\ }\bibinfo
  {organization} {CERN}\ (\bibinfo  {publisher} {CERN},\ \bibinfo {address}
  {Geneva},\ \bibinfo {year} {2018})\BibitemShut {NoStop}%
\bibitem [{\citenamefont {Adamczyk}\ \emph {et~al.}(2015)\citenamefont
  {Adamczyk} \emph {et~al.}}]{Adamczyk:2015obl}%
  \BibitemOpen
  \bibfield  {author} {\bibinfo {author} {\bibfnamefont {L.}~\bibnamefont
  {Adamczyk}} \emph {et~al.} (\bibinfo {collaboration} {STAR}),\ }\href
  {\doibase 10.1103/PhysRevLett.115.222301} {\bibfield  {journal} {\bibinfo
  {journal} {Phys. Rev. Lett.}\ }\textbf {\bibinfo {volume} {115}},\ \bibinfo
  {pages} {222301} (\bibinfo {year} {2015})},\ \Eprint
  {http://arxiv.org/abs/1505.07812} {arXiv:1505.07812 [nucl-ex]} \BibitemShut
  {NoStop}%
\bibitem [{\citenamefont {Goldschmidt}\ \emph {et~al.}(2015)\citenamefont
  {Goldschmidt}, \citenamefont {Qiu}, \citenamefont {Shen},\ and\ \citenamefont
  {Heinz}}]{Goldschmidt:2015kpa}%
  \BibitemOpen
  \bibfield  {author} {\bibinfo {author} {\bibfnamefont {A.}~\bibnamefont
  {Goldschmidt}}, \bibinfo {author} {\bibfnamefont {Z.}~\bibnamefont {Qiu}},
  \bibinfo {author} {\bibfnamefont {C.}~\bibnamefont {Shen}}, \ and\ \bibinfo
  {author} {\bibfnamefont {U.}~\bibnamefont {Heinz}},\ }\href {\doibase
  10.1103/PhysRevC.92.044903} {\bibfield  {journal} {\bibinfo  {journal} {Phys.
  Rev.}\ }\textbf {\bibinfo {volume} {C92}},\ \bibinfo {pages} {044903}
  (\bibinfo {year} {2015})},\ \Eprint {http://arxiv.org/abs/1507.03910}
  {arXiv:1507.03910 [nucl-th]} \BibitemShut {NoStop}%
\bibitem [{\citenamefont {Dasgupta}\ \emph {et~al.}(2017)\citenamefont
  {Dasgupta}, \citenamefont {Chatterjee},\ and\ \citenamefont
  {Srivastava}}]{Dasgupta:2016qkq}%
  \BibitemOpen
  \bibfield  {author} {\bibinfo {author} {\bibfnamefont {P.}~\bibnamefont
  {Dasgupta}}, \bibinfo {author} {\bibfnamefont {R.}~\bibnamefont
  {Chatterjee}}, \ and\ \bibinfo {author} {\bibfnamefont {D.~K.}\ \bibnamefont
  {Srivastava}},\ }\href {\doibase 10.1103/PhysRevC.95.064907} {\bibfield
  {journal} {\bibinfo  {journal} {Phys. Rev.}\ }\textbf {\bibinfo {volume}
  {C95}},\ \bibinfo {pages} {064907} (\bibinfo {year} {2017})},\ \Eprint
  {http://arxiv.org/abs/1609.01949} {arXiv:1609.01949 [nucl-th]} \BibitemShut
  {NoStop}%
\bibitem [{\citenamefont {Giacalone}(2019{\natexlab{a}})}]{Giacalone:2018apa}%
  \BibitemOpen
  \bibfield  {author} {\bibinfo {author} {\bibfnamefont {G.}~\bibnamefont
  {Giacalone}},\ }\href {\doibase 10.1103/PhysRevC.99.024910} {\bibfield
  {journal} {\bibinfo  {journal} {Phys. Rev.}\ }\textbf {\bibinfo {volume}
  {C99}},\ \bibinfo {pages} {024910} (\bibinfo {year} {2019}{\natexlab{a}})},\
  \Eprint {http://arxiv.org/abs/1811.03959} {arXiv:1811.03959 [nucl-th]}
  \BibitemShut {NoStop}%
\bibitem [{\citenamefont {Lim}\ \emph {et~al.}(2019)\citenamefont {Lim},
  \citenamefont {Hu}, \citenamefont {Belmont}, \citenamefont {Hill},
  \citenamefont {Nagle},\ and\ \citenamefont {Perepelitsa}}]{Lim:2019cys}%
  \BibitemOpen
  \bibfield  {author} {\bibinfo {author} {\bibfnamefont {S.~H.}\ \bibnamefont
  {Lim}}, \bibinfo {author} {\bibfnamefont {Q.}~\bibnamefont {Hu}}, \bibinfo
  {author} {\bibfnamefont {R.}~\bibnamefont {Belmont}}, \bibinfo {author}
  {\bibfnamefont {K.~K.}\ \bibnamefont {Hill}}, \bibinfo {author}
  {\bibfnamefont {J.~L.}\ \bibnamefont {Nagle}}, \ and\ \bibinfo {author}
  {\bibfnamefont {D.~V.}\ \bibnamefont {Perepelitsa}},\ }\href {\doibase
  10.1103/PhysRevC.100.024908} {\bibfield  {journal} {\bibinfo  {journal}
  {Phys. Rev.}\ }\textbf {\bibinfo {volume} {C100}},\ \bibinfo {pages} {024908}
  (\bibinfo {year} {2019})},\ \Eprint {http://arxiv.org/abs/1902.11290}
  {arXiv:1902.11290 [nucl-th]} \BibitemShut {NoStop}%
\bibitem [{\citenamefont {Citron}\ \emph {et~al.}(2019)\citenamefont {Citron}
  \emph {et~al.}}]{Citron:2018lsq}%
  \BibitemOpen
  \bibfield  {author} {\bibinfo {author} {\bibfnamefont {Z.}~\bibnamefont
  {Citron}} \emph {et~al.},\ }in\ \href {\doibase 10.23731/CYRM-2019-007.1159}
  {\emph {\bibinfo {booktitle} {Report on the Physics at the HL-LHC,and
  Perspectives for the HE-LHC}}},\ \bibinfo {editor} {edited by\ \bibinfo
  {editor} {\bibfnamefont {A.}~\bibnamefont {Dainese}}, \bibinfo {editor}
  {\bibfnamefont {M.}~\bibnamefont {Mangano}}, \bibinfo {editor} {\bibfnamefont
  {A.~B.}\ \bibnamefont {Meyer}}, \bibinfo {editor} {\bibfnamefont
  {A.}~\bibnamefont {Nisati}}, \bibinfo {editor} {\bibfnamefont
  {G.}~\bibnamefont {Salam}}, \ and\ \bibinfo {editor} {\bibfnamefont {M.~A.}\
  \bibnamefont {Vesterinen}}}\ (\bibinfo {year} {2019})\ pp.\ \bibinfo {pages}
  {1159--1410},\ \Eprint {http://arxiv.org/abs/1812.06772} {arXiv:1812.06772
  [hep-ph]} \BibitemShut {NoStop}%
\bibitem [{\citenamefont {Rybczyński}\ and\ \citenamefont
  {Broniowski}(2019)}]{Rybczynski:2019adt}%
  \BibitemOpen
  \bibfield  {author} {\bibinfo {author} {\bibfnamefont {M.}~\bibnamefont
  {Rybczyński}}\ and\ \bibinfo {author} {\bibfnamefont {W.}~\bibnamefont
  {Broniowski}},\ }\href {\doibase 10.1103/PhysRevC.100.064912} {\bibfield
  {journal} {\bibinfo  {journal} {Phys. Rev.}\ }\textbf {\bibinfo {volume}
  {C100}},\ \bibinfo {pages} {064912} (\bibinfo {year} {2019})},\ \Eprint
  {http://arxiv.org/abs/1910.09489} {arXiv:1910.09489 [hep-ph]} \BibitemShut
  {NoStop}%
\bibitem [{\citenamefont {Noronha-Hostler}\ \emph {et~al.}(2019)\citenamefont
  {Noronha-Hostler}, \citenamefont {Paladino}, \citenamefont {Rao},
  \citenamefont {Sievert},\ and\ \citenamefont
  {Wertepny}}]{Noronha-Hostler:2019ytn}%
  \BibitemOpen
  \bibfield  {author} {\bibinfo {author} {\bibfnamefont {J.}~\bibnamefont
  {Noronha-Hostler}}, \bibinfo {author} {\bibfnamefont {N.}~\bibnamefont
  {Paladino}}, \bibinfo {author} {\bibfnamefont {S.}~\bibnamefont {Rao}},
  \bibinfo {author} {\bibfnamefont {M.~D.}\ \bibnamefont {Sievert}}, \ and\
  \bibinfo {author} {\bibfnamefont {D.~E.}\ \bibnamefont {Wertepny}},\
  }\href@noop {} {\  (\bibinfo {year} {2019})},\ \Eprint
  {http://arxiv.org/abs/1905.13323} {arXiv:1905.13323 [hep-ph]} \BibitemShut
  {NoStop}%
\bibitem [{\citenamefont {Bozek}\ and\ \citenamefont
  {Broniowski}(2018)}]{Bozek:2018xzy}%
  \BibitemOpen
  \bibfield  {author} {\bibinfo {author} {\bibfnamefont {P.}~\bibnamefont
  {Bozek}}\ and\ \bibinfo {author} {\bibfnamefont {W.}~\bibnamefont
  {Broniowski}},\ }\href {\doibase 10.1103/PhysRevLett.121.202301} {\bibfield
  {journal} {\bibinfo  {journal} {Phys. Rev. Lett.}\ }\textbf {\bibinfo
  {volume} {121}},\ \bibinfo {pages} {202301} (\bibinfo {year} {2018})},\
  \Eprint {http://arxiv.org/abs/1808.09840} {arXiv:1808.09840 [nucl-th]}
  \BibitemShut {NoStop}%
\bibitem [{\citenamefont {Broniowski}\ and\ \citenamefont
  {Bożek}(2019)}]{Broniowski:2019kjo}%
  \BibitemOpen
  \bibfield  {author} {\bibinfo {author} {\bibfnamefont {W.}~\bibnamefont
  {Broniowski}}\ and\ \bibinfo {author} {\bibfnamefont {P.}~\bibnamefont
  {Bożek}},\ }\href@noop {} {\  (\bibinfo {year} {2019})},\ \Eprint
  {http://arxiv.org/abs/1906.09045} {arXiv:1906.09045 [nucl-th]} \BibitemShut
  {NoStop}%
\bibitem [{\citenamefont {Giacalone}(2019{\natexlab{b}})}]{Giacalone:2019pca}%
  \BibitemOpen
  \bibfield  {author} {\bibinfo {author} {\bibfnamefont {G.}~\bibnamefont
  {Giacalone}},\ }\href@noop {} {\  (\bibinfo {year} {2019}{\natexlab{b}})},\
  \Eprint {http://arxiv.org/abs/1910.04673} {arXiv:1910.04673 [nucl-th]}
  \BibitemShut {NoStop}%
\bibitem [{\citenamefont {Giacalone}(2020)}]{Giacalone:2020awm}%
  \BibitemOpen
  \bibfield  {author} {\bibinfo {author} {\bibfnamefont {G.}~\bibnamefont
  {Giacalone}},\ }\href@noop {} {\  (\bibinfo {year} {2020})},\ \Eprint
  {http://arxiv.org/abs/2004.14463} {arXiv:2004.14463 [nucl-th]} \BibitemShut
  {NoStop}%
\bibitem [{\citenamefont {Luzum}\ and\ \citenamefont
  {Ollitrault}(2013)}]{Luzum:2012wu}%
  \BibitemOpen
  \bibfield  {author} {\bibinfo {author} {\bibfnamefont {M.}~\bibnamefont
  {Luzum}}\ and\ \bibinfo {author} {\bibfnamefont {J.-Y.}\ \bibnamefont
  {Ollitrault}},\ }\bibfield  {booktitle} {\emph {\bibinfo {booktitle}
  {{Proceedings, 23rd International Conference on Ultrarelativistic
  Nucleus-Nucleus Collisions : Quark Matter 2012 (QM 2012): Washington, DC,
  USA, August 13-18, 2012}}},\ }\href {\doibase
  10.1016/j.nuclphysa.2013.02.028} {\bibfield  {journal} {\bibinfo  {journal}
  {Nucl. Phys.}\ }\textbf {\bibinfo {volume} {A904-905}},\ \bibinfo {pages}
  {377c} (\bibinfo {year} {2013})},\ \Eprint {http://arxiv.org/abs/1210.6010}
  {arXiv:1210.6010 [nucl-th]} \BibitemShut {NoStop}%
\bibitem [{\citenamefont {Collaboration}(2012)}]{CMS:2012xxa}%
  \BibitemOpen
  \bibfield  {author} {\bibinfo {author} {\bibfnamefont {C.}~\bibnamefont
  {Collaboration}} (\bibinfo {collaboration} {CMS}),\ }\href@noop {} {\
  (\bibinfo {year} {2012})}\BibitemShut {NoStop}%
\bibitem [{\citenamefont {Shen}\ \emph {et~al.}(2015)\citenamefont {Shen},
  \citenamefont {Qiu},\ and\ \citenamefont {Heinz}}]{Shen:2015qta}%
  \BibitemOpen
  \bibfield  {author} {\bibinfo {author} {\bibfnamefont {C.}~\bibnamefont
  {Shen}}, \bibinfo {author} {\bibfnamefont {Z.}~\bibnamefont {Qiu}}, \ and\
  \bibinfo {author} {\bibfnamefont {U.}~\bibnamefont {Heinz}},\ }\href
  {\doibase 10.1103/PhysRevC.92.014901} {\bibfield  {journal} {\bibinfo
  {journal} {Phys. Rev.}\ }\textbf {\bibinfo {volume} {C92}},\ \bibinfo {pages}
  {014901} (\bibinfo {year} {2015})},\ \Eprint
  {http://arxiv.org/abs/1502.04636} {arXiv:1502.04636 [nucl-th]} \BibitemShut
  {NoStop}%
\bibitem [{\citenamefont {Rose}\ \emph {et~al.}(2014)\citenamefont {Rose},
  \citenamefont {Paquet}, \citenamefont {Denicol}, \citenamefont {Luzum},
  \citenamefont {Schenke}, \citenamefont {Jeon},\ and\ \citenamefont
  {Gale}}]{Rose:2014fba}%
  \BibitemOpen
  \bibfield  {author} {\bibinfo {author} {\bibfnamefont {J.-B.}\ \bibnamefont
  {Rose}}, \bibinfo {author} {\bibfnamefont {J.-F.}\ \bibnamefont {Paquet}},
  \bibinfo {author} {\bibfnamefont {G.~S.}\ \bibnamefont {Denicol}}, \bibinfo
  {author} {\bibfnamefont {M.}~\bibnamefont {Luzum}}, \bibinfo {author}
  {\bibfnamefont {B.}~\bibnamefont {Schenke}}, \bibinfo {author} {\bibfnamefont
  {S.}~\bibnamefont {Jeon}}, \ and\ \bibinfo {author} {\bibfnamefont
  {C.}~\bibnamefont {Gale}},\ }\bibfield  {booktitle} {\emph {\bibinfo
  {booktitle} {{Proceedings, 24th International Conference on
  Ultra-Relativistic Nucleus-Nucleus Collisions (Quark Matter 2014): Darmstadt,
  Germany, May 19-24, 2014}}},\ }\href {\doibase
  10.1016/j.nuclphysa.2014.09.044} {\bibfield  {journal} {\bibinfo  {journal}
  {Nucl. Phys.}\ }\textbf {\bibinfo {volume} {A931}},\ \bibinfo {pages} {926}
  (\bibinfo {year} {2014})},\ \Eprint {http://arxiv.org/abs/1408.0024}
  {arXiv:1408.0024 [nucl-th]} \BibitemShut {NoStop}%
\bibitem [{\citenamefont {Gelis}\ \emph {et~al.}(2019)\citenamefont {Gelis},
  \citenamefont {Giacalone}, \citenamefont {Guerrero-Rodríguez}, \citenamefont
  {Marquet},\ and\ \citenamefont {Ollitrault}}]{Gelis:2019vzt}%
  \BibitemOpen
  \bibfield  {author} {\bibinfo {author} {\bibfnamefont {F.}~\bibnamefont
  {Gelis}}, \bibinfo {author} {\bibfnamefont {G.}~\bibnamefont {Giacalone}},
  \bibinfo {author} {\bibfnamefont {P.}~\bibnamefont {Guerrero-Rodríguez}},
  \bibinfo {author} {\bibfnamefont {C.}~\bibnamefont {Marquet}}, \ and\
  \bibinfo {author} {\bibfnamefont {J.-Y.}\ \bibnamefont {Ollitrault}},\
  }\href@noop {} {\  (\bibinfo {year} {2019})},\ \Eprint
  {http://arxiv.org/abs/1907.10948} {arXiv:1907.10948 [nucl-th]} \BibitemShut
  {NoStop}%
\bibitem [{\citenamefont {Giacalone}\ \emph {et~al.}(2017)\citenamefont
  {Giacalone}, \citenamefont {Noronha-Hostler},\ and\ \citenamefont
  {Ollitrault}}]{Giacalone:2017uqx}%
  \BibitemOpen
  \bibfield  {author} {\bibinfo {author} {\bibfnamefont {G.}~\bibnamefont
  {Giacalone}}, \bibinfo {author} {\bibfnamefont {J.}~\bibnamefont
  {Noronha-Hostler}}, \ and\ \bibinfo {author} {\bibfnamefont {J.-Y.}\
  \bibnamefont {Ollitrault}},\ }\href {\doibase 10.1103/PhysRevC.95.054910}
  {\bibfield  {journal} {\bibinfo  {journal} {Phys. Rev.}\ }\textbf {\bibinfo
  {volume} {C95}},\ \bibinfo {pages} {054910} (\bibinfo {year} {2017})},\
  \Eprint {http://arxiv.org/abs/1702.01730} {arXiv:1702.01730 [nucl-th]}
  \BibitemShut {NoStop}%
\bibitem [{\citenamefont {Chatrchyan}\ \emph {et~al.}(2014)\citenamefont
  {Chatrchyan} \emph {et~al.}}]{Chatrchyan:2013kba}%
  \BibitemOpen
  \bibfield  {author} {\bibinfo {author} {\bibfnamefont {S.}~\bibnamefont
  {Chatrchyan}} \emph {et~al.} (\bibinfo {collaboration} {CMS}),\ }\href
  {\doibase 10.1103/PhysRevC.89.044906} {\bibfield  {journal} {\bibinfo
  {journal} {Phys. Rev.}\ }\textbf {\bibinfo {volume} {C89}},\ \bibinfo {pages}
  {044906} (\bibinfo {year} {2014})},\ \Eprint {http://arxiv.org/abs/1310.8651}
  {arXiv:1310.8651 [nucl-ex]} \BibitemShut {NoStop}%
\bibitem [{\citenamefont {Aamodt}\ \emph {et~al.}(2011)\citenamefont {Aamodt}
  \emph {et~al.}}]{ALICE:2011ab}%
  \BibitemOpen
  \bibfield  {author} {\bibinfo {author} {\bibfnamefont {K.}~\bibnamefont
  {Aamodt}} \emph {et~al.} (\bibinfo {collaboration} {ALICE}),\ }\href
  {\doibase 10.1103/PhysRevLett.107.032301} {\bibfield  {journal} {\bibinfo
  {journal} {Phys. Rev. Lett.}\ }\textbf {\bibinfo {volume} {107}},\ \bibinfo
  {pages} {032301} (\bibinfo {year} {2011})},\ \Eprint
  {http://arxiv.org/abs/1105.3865} {arXiv:1105.3865 [nucl-ex]} \BibitemShut
  {NoStop}%
\bibitem [{\citenamefont {Aad}\ \emph {et~al.}(2014)\citenamefont {Aad} \emph
  {et~al.}}]{Aad:2014vba}%
  \BibitemOpen
  \bibfield  {author} {\bibinfo {author} {\bibfnamefont {G.}~\bibnamefont
  {Aad}} \emph {et~al.} (\bibinfo {collaboration} {ATLAS}),\ }\href {\doibase
  10.1140/epjc/s10052-014-3157-z} {\bibfield  {journal} {\bibinfo  {journal}
  {Eur. Phys. J.}\ }\textbf {\bibinfo {volume} {C74}},\ \bibinfo {pages} {3157}
  (\bibinfo {year} {2014})},\ \Eprint {http://arxiv.org/abs/1408.4342}
  {arXiv:1408.4342 [hep-ex]} \BibitemShut {NoStop}%
\bibitem [{\citenamefont {Noronha-Hostler}\ \emph
  {et~al.}(2016{\natexlab{c}})\citenamefont {Noronha-Hostler}, \citenamefont
  {Yan}, \citenamefont {Gardim},\ and\ \citenamefont
  {Ollitrault}}]{Noronha-Hostler:2015dbi}%
  \BibitemOpen
  \bibfield  {author} {\bibinfo {author} {\bibfnamefont {J.}~\bibnamefont
  {Noronha-Hostler}}, \bibinfo {author} {\bibfnamefont {L.}~\bibnamefont
  {Yan}}, \bibinfo {author} {\bibfnamefont {F.~G.}\ \bibnamefont {Gardim}}, \
  and\ \bibinfo {author} {\bibfnamefont {J.-Y.}\ \bibnamefont {Ollitrault}},\
  }\href {\doibase 10.1103/PhysRevC.93.014909} {\bibfield  {journal} {\bibinfo
  {journal} {Phys. Rev.}\ }\textbf {\bibinfo {volume} {C93}},\ \bibinfo {pages}
  {014909} (\bibinfo {year} {2016}{\natexlab{c}})},\ \Eprint
  {http://arxiv.org/abs/1511.03896} {arXiv:1511.03896 [nucl-th]} \BibitemShut
  {NoStop}%
\bibitem [{\citenamefont {Sievert}\ and\ \citenamefont
  {Noronha-Hostler}(2019)}]{Sievert:2019zjr}%
  \BibitemOpen
  \bibfield  {author} {\bibinfo {author} {\bibfnamefont {M.~D.}\ \bibnamefont
  {Sievert}}\ and\ \bibinfo {author} {\bibfnamefont {J.}~\bibnamefont
  {Noronha-Hostler}},\ }\href {\doibase 10.1103/PhysRevC.100.024904} {\bibfield
   {journal} {\bibinfo  {journal} {Phys. Rev.}\ }\textbf {\bibinfo {volume}
  {C100}},\ \bibinfo {pages} {024904} (\bibinfo {year} {2019})},\ \Eprint
  {http://arxiv.org/abs/1901.01319} {arXiv:1901.01319 [nucl-th]} \BibitemShut
  {NoStop}%
\bibitem [{\citenamefont {Rao}\ \emph {et~al.}(2019)\citenamefont {Rao},
  \citenamefont {Sievert},\ and\ \citenamefont
  {Noronha-Hostler}}]{Rao:2019vgy}%
  \BibitemOpen
  \bibfield  {author} {\bibinfo {author} {\bibfnamefont {S.}~\bibnamefont
  {Rao}}, \bibinfo {author} {\bibfnamefont {M.}~\bibnamefont {Sievert}}, \ and\
  \bibinfo {author} {\bibfnamefont {J.}~\bibnamefont {Noronha-Hostler}},\
  }\href@noop {} {\  (\bibinfo {year} {2019})},\ \Eprint
  {http://arxiv.org/abs/1910.03677} {arXiv:1910.03677 [nucl-th]} \BibitemShut
  {NoStop}%
\bibitem [{\citenamefont {Möller}\ \emph {et~al.}(2016)\citenamefont
  {Möller}, \citenamefont {Sierk}, \citenamefont {Ichikawa},\ and\
  \citenamefont {Sagawa}}]{Moller:2015fba}%
  \BibitemOpen
  \bibfield  {author} {\bibinfo {author} {\bibfnamefont {P.}~\bibnamefont
  {Möller}}, \bibinfo {author} {\bibfnamefont {A.~J.}\ \bibnamefont {Sierk}},
  \bibinfo {author} {\bibfnamefont {T.}~\bibnamefont {Ichikawa}}, \ and\
  \bibinfo {author} {\bibfnamefont {H.}~\bibnamefont {Sagawa}},\ }\href
  {\doibase 10.1016/j.adt.2015.10.002} {\bibfield  {journal} {\bibinfo
  {journal} {Atom. Data Nucl. Data Tabl.}\ }\textbf {\bibinfo {volume}
  {109-110}},\ \bibinfo {pages} {1} (\bibinfo {year} {2016})},\ \Eprint
  {http://arxiv.org/abs/1508.06294} {arXiv:1508.06294 [nucl-th]} \BibitemShut
  {NoStop}%
\bibitem [{\citenamefont {Robledo}\ and\ \citenamefont
  {Bertsch}(2011)}]{Robledo:2011nf}%
  \BibitemOpen
  \bibfield  {author} {\bibinfo {author} {\bibfnamefont {L.~M.}\ \bibnamefont
  {Robledo}}\ and\ \bibinfo {author} {\bibfnamefont {G.~F.}\ \bibnamefont
  {Bertsch}},\ }\href {\doibase 10.1103/PhysRevC.84.054302} {\bibfield
  {journal} {\bibinfo  {journal} {Phys. Rev.}\ }\textbf {\bibinfo {volume}
  {C84}},\ \bibinfo {pages} {054302} (\bibinfo {year} {2011})},\ \Eprint
  {http://arxiv.org/abs/1107.3581} {arXiv:1107.3581 [nucl-th]} \BibitemShut
  {NoStop}%
\bibitem [{\citenamefont {Tarbert}\ \emph {et~al.}(2014)\citenamefont {Tarbert}
  \emph {et~al.}}]{Tarbert:2013jze}%
  \BibitemOpen
  \bibfield  {author} {\bibinfo {author} {\bibfnamefont {C.~M.}\ \bibnamefont
  {Tarbert}} \emph {et~al.},\ }\href {\doibase 10.1103/PhysRevLett.112.242502}
  {\bibfield  {journal} {\bibinfo  {journal} {Phys. Rev. Lett.}\ }\textbf
  {\bibinfo {volume} {112}},\ \bibinfo {pages} {242502} (\bibinfo {year}
  {2014})},\ \Eprint {http://arxiv.org/abs/1311.0168} {arXiv:1311.0168
  [nucl-ex]} \BibitemShut {NoStop}%
\bibitem [{\citenamefont {Abrahamyan}\ \emph {et~al.}(2012)\citenamefont
  {Abrahamyan} \emph {et~al.}}]{Abrahamyan:2012gp}%
  \BibitemOpen
  \bibfield  {author} {\bibinfo {author} {\bibfnamefont {S.}~\bibnamefont
  {Abrahamyan}} \emph {et~al.},\ }\href {\doibase
  10.1103/PhysRevLett.108.112502} {\bibfield  {journal} {\bibinfo  {journal}
  {Phys. Rev. Lett.}\ }\textbf {\bibinfo {volume} {108}},\ \bibinfo {pages}
  {112502} (\bibinfo {year} {2012})},\ \Eprint {http://arxiv.org/abs/1201.2568}
  {arXiv:1201.2568 [nucl-ex]} \BibitemShut {NoStop}%
\bibitem [{\citenamefont {Horowitz}\ and\ \citenamefont
  {Piekarewicz}(2001)}]{Horowitz:2000xj}%
  \BibitemOpen
  \bibfield  {author} {\bibinfo {author} {\bibfnamefont {C.~J.}\ \bibnamefont
  {Horowitz}}\ and\ \bibinfo {author} {\bibfnamefont {J.}~\bibnamefont
  {Piekarewicz}},\ }\href {\doibase 10.1103/PhysRevLett.86.5647} {\bibfield
  {journal} {\bibinfo  {journal} {Phys. Rev. Lett.}\ }\textbf {\bibinfo
  {volume} {86}},\ \bibinfo {pages} {5647} (\bibinfo {year} {2001})},\ \Eprint
  {http://arxiv.org/abs/astro-ph/0010227} {arXiv:astro-ph/0010227 [astro-ph]}
  \BibitemShut {NoStop}%
\bibitem [{\citenamefont {Fattoyev}\ and\ \citenamefont
  {Piekarewicz}(2012)}]{Fattoyev:2012rm}%
  \BibitemOpen
  \bibfield  {author} {\bibinfo {author} {\bibfnamefont {F.~J.}\ \bibnamefont
  {Fattoyev}}\ and\ \bibinfo {author} {\bibfnamefont {J.}~\bibnamefont
  {Piekarewicz}},\ }\href {\doibase 10.1103/PhysRevC.86.015802} {\bibfield
  {journal} {\bibinfo  {journal} {Phys. Rev.}\ }\textbf {\bibinfo {volume}
  {C86}},\ \bibinfo {pages} {015802} (\bibinfo {year} {2012})},\ \Eprint
  {http://arxiv.org/abs/1203.4006} {arXiv:1203.4006 [nucl-th]} \BibitemShut
  {NoStop}%
\bibitem [{\citenamefont {Fattoyev}\ \emph {et~al.}(2018)\citenamefont
  {Fattoyev}, \citenamefont {Piekarewicz},\ and\ \citenamefont
  {Horowitz}}]{Fattoyev:2017jql}%
  \BibitemOpen
  \bibfield  {author} {\bibinfo {author} {\bibfnamefont {F.~J.}\ \bibnamefont
  {Fattoyev}}, \bibinfo {author} {\bibfnamefont {J.}~\bibnamefont
  {Piekarewicz}}, \ and\ \bibinfo {author} {\bibfnamefont {C.~J.}\ \bibnamefont
  {Horowitz}},\ }\href {\doibase 10.1103/PhysRevLett.120.172702} {\bibfield
  {journal} {\bibinfo  {journal} {Phys. Rev. Lett.}\ }\textbf {\bibinfo
  {volume} {120}},\ \bibinfo {pages} {172702} (\bibinfo {year} {2018})},\
  \Eprint {http://arxiv.org/abs/1711.06615} {arXiv:1711.06615 [nucl-th]}
  \BibitemShut {NoStop}%
\bibitem [{\citenamefont {Aaboud}\ \emph {et~al.}(2020)\citenamefont {Aaboud}
  \emph {et~al.}}]{Aaboud:2019sma}%
  \BibitemOpen
  \bibfield  {author} {\bibinfo {author} {\bibfnamefont {M.}~\bibnamefont
  {Aaboud}} \emph {et~al.} (\bibinfo {collaboration} {ATLAS}),\ }\href
  {\doibase 10.1007/JHEP01(2020)051} {\bibfield  {journal} {\bibinfo  {journal}
  {JHEP}\ }\textbf {\bibinfo {volume} {01}},\ \bibinfo {pages} {051} (\bibinfo
  {year} {2020})},\ \Eprint {http://arxiv.org/abs/1904.04808} {arXiv:1904.04808
  [nucl-ex]} \BibitemShut {NoStop}%
\bibitem [{\citenamefont {Moreland}\ \emph {et~al.}(2015)\citenamefont
  {Moreland}, \citenamefont {Bernhard},\ and\ \citenamefont
  {Bass}}]{Moreland:2014oya}%
  \BibitemOpen
  \bibfield  {author} {\bibinfo {author} {\bibfnamefont {J.~S.}\ \bibnamefont
  {Moreland}}, \bibinfo {author} {\bibfnamefont {J.~E.}\ \bibnamefont
  {Bernhard}}, \ and\ \bibinfo {author} {\bibfnamefont {S.~A.}\ \bibnamefont
  {Bass}},\ }\href {\doibase 10.1103/PhysRevC.92.011901} {\bibfield  {journal}
  {\bibinfo  {journal} {Phys. Rev.}\ }\textbf {\bibinfo {volume} {C92}},\
  \bibinfo {pages} {011901} (\bibinfo {year} {2015})},\ \Eprint
  {http://arxiv.org/abs/1412.4708} {arXiv:1412.4708 [nucl-th]} \BibitemShut
  {NoStop}%
\bibitem [{\citenamefont {Lappi}(2006)}]{Lappi:2006hq}%
  \BibitemOpen
  \bibfield  {author} {\bibinfo {author} {\bibfnamefont {T.}~\bibnamefont
  {Lappi}},\ }\href {\doibase 10.1016/j.physletb.2006.10.017} {\bibfield
  {journal} {\bibinfo  {journal} {Phys. Lett. B}\ }\textbf {\bibinfo {volume}
  {643}},\ \bibinfo {pages} {11} (\bibinfo {year} {2006})},\ \Eprint
  {http://arxiv.org/abs/hep-ph/0606207} {arXiv:hep-ph/0606207} \BibitemShut
  {NoStop}%
\bibitem [{\citenamefont {Romatschke}\ and\ \citenamefont
  {Romatschke}(2019)}]{Romatschke:2017ejr}%
  \BibitemOpen
  \bibfield  {author} {\bibinfo {author} {\bibfnamefont {P.}~\bibnamefont
  {Romatschke}}\ and\ \bibinfo {author} {\bibfnamefont {U.}~\bibnamefont
  {Romatschke}},\ }\href {\doibase 10.1017/9781108651998} {\emph {\bibinfo
  {title} {{Relativistic Fluid Dynamics In and Out of Equilibrium}}}},\
  Cambridge Monographs on Mathematical Physics\ (\bibinfo  {publisher}
  {Cambridge University Press},\ \bibinfo {year} {2019})\ \Eprint
  {http://arxiv.org/abs/1712.05815} {arXiv:1712.05815 [nucl-th]} \BibitemShut
  {NoStop}%
\bibitem [{\citenamefont {Chen}\ \emph {et~al.}(2015)\citenamefont {Chen},
  \citenamefont {Fries}, \citenamefont {Kapusta},\ and\ \citenamefont
  {Li}}]{Chen:2015wia}%
  \BibitemOpen
  \bibfield  {author} {\bibinfo {author} {\bibfnamefont {G.}~\bibnamefont
  {Chen}}, \bibinfo {author} {\bibfnamefont {R.~J.}\ \bibnamefont {Fries}},
  \bibinfo {author} {\bibfnamefont {J.~I.}\ \bibnamefont {Kapusta}}, \ and\
  \bibinfo {author} {\bibfnamefont {Y.}~\bibnamefont {Li}},\ }\href {\doibase
  10.1103/PhysRevC.92.064912} {\bibfield  {journal} {\bibinfo  {journal} {Phys.
  Rev. C}\ }\textbf {\bibinfo {volume} {92}},\ \bibinfo {pages} {064912}
  (\bibinfo {year} {2015})},\ \Eprint {http://arxiv.org/abs/1507.03524}
  {arXiv:1507.03524 [nucl-th]} \BibitemShut {NoStop}%
\bibitem [{\citenamefont {Acharya}\ \emph
  {et~al.}(2018{\natexlab{b}})\citenamefont {Acharya} \emph
  {et~al.}}]{Acharya:2018lmh}%
  \BibitemOpen
  \bibfield  {author} {\bibinfo {author} {\bibfnamefont {S.}~\bibnamefont
  {Acharya}} \emph {et~al.} (\bibinfo {collaboration} {ALICE}),\ }\href
  {\doibase 10.1007/JHEP07(2018)103} {\bibfield  {journal} {\bibinfo  {journal}
  {JHEP}\ }\textbf {\bibinfo {volume} {07}},\ \bibinfo {pages} {103} (\bibinfo
  {year} {2018}{\natexlab{b}})},\ \Eprint {http://arxiv.org/abs/1804.02944}
  {arXiv:1804.02944 [nucl-ex]} \BibitemShut {NoStop}%
\bibitem [{\citenamefont {Noronha-Hostler}\ \emph {et~al.}(2013)\citenamefont
  {Noronha-Hostler}, \citenamefont {Denicol}, \citenamefont {Noronha},
  \citenamefont {Andrade},\ and\ \citenamefont
  {Grassi}}]{Noronha-Hostler:2013gga}%
  \BibitemOpen
  \bibfield  {author} {\bibinfo {author} {\bibfnamefont {J.}~\bibnamefont
  {Noronha-Hostler}}, \bibinfo {author} {\bibfnamefont {G.~S.}\ \bibnamefont
  {Denicol}}, \bibinfo {author} {\bibfnamefont {J.}~\bibnamefont {Noronha}},
  \bibinfo {author} {\bibfnamefont {R.~P.~G.}\ \bibnamefont {Andrade}}, \ and\
  \bibinfo {author} {\bibfnamefont {F.}~\bibnamefont {Grassi}},\ }\href
  {\doibase 10.1103/PhysRevC.88.044916} {\bibfield  {journal} {\bibinfo
  {journal} {Phys. Rev. C}\ }\textbf {\bibinfo {volume} {88}},\ \bibinfo
  {pages} {044916} (\bibinfo {year} {2013})},\ \Eprint
  {http://arxiv.org/abs/1305.1981} {arXiv:1305.1981 [nucl-th]} \BibitemShut
  {NoStop}%
\bibitem [{\citenamefont {Noronha-Hostler}\ \emph {et~al.}(2014)\citenamefont
  {Noronha-Hostler}, \citenamefont {Noronha},\ and\ \citenamefont
  {Grassi}}]{Noronha-Hostler:2014dqa}%
  \BibitemOpen
  \bibfield  {author} {\bibinfo {author} {\bibfnamefont {J.}~\bibnamefont
  {Noronha-Hostler}}, \bibinfo {author} {\bibfnamefont {J.}~\bibnamefont
  {Noronha}}, \ and\ \bibinfo {author} {\bibfnamefont {F.}~\bibnamefont
  {Grassi}},\ }\href {\doibase 10.1103/PhysRevC.90.034907} {\bibfield
  {journal} {\bibinfo  {journal} {Phys. Rev. C}\ }\textbf {\bibinfo {volume}
  {90}},\ \bibinfo {pages} {034907} (\bibinfo {year} {2014})},\ \Eprint
  {http://arxiv.org/abs/1406.3333} {arXiv:1406.3333 [nucl-th]} \BibitemShut
  {NoStop}%
\bibitem [{\citenamefont {Sollfrank}\ \emph {et~al.}(1991)\citenamefont
  {Sollfrank}, \citenamefont {Koch},\ and\ \citenamefont
  {Heinz}}]{Sollfrank:1991xm}%
  \BibitemOpen
  \bibfield  {author} {\bibinfo {author} {\bibfnamefont {J.}~\bibnamefont
  {Sollfrank}}, \bibinfo {author} {\bibfnamefont {P.}~\bibnamefont {Koch}}, \
  and\ \bibinfo {author} {\bibfnamefont {U.~W.}\ \bibnamefont {Heinz}},\ }\href
  {\doibase 10.1007/BF01562334} {\bibfield  {journal} {\bibinfo  {journal} {Z.
  Phys. C}\ }\textbf {\bibinfo {volume} {52}},\ \bibinfo {pages} {593}
  (\bibinfo {year} {1991})}\BibitemShut {NoStop}%
\bibitem [{\citenamefont {Wiedemann}\ and\ \citenamefont
  {Heinz}(1997)}]{Wiedemann:1996ig}%
  \BibitemOpen
  \bibfield  {author} {\bibinfo {author} {\bibfnamefont {U.~A.}\ \bibnamefont
  {Wiedemann}}\ and\ \bibinfo {author} {\bibfnamefont {U.~W.}\ \bibnamefont
  {Heinz}},\ }\href {\doibase 10.1103/PhysRevC.56.3265} {\bibfield  {journal}
  {\bibinfo  {journal} {Phys. Rev. C}\ }\textbf {\bibinfo {volume} {56}},\
  \bibinfo {pages} {3265} (\bibinfo {year} {1997})},\ \Eprint
  {http://arxiv.org/abs/nucl-th/9611031} {arXiv:nucl-th/9611031} \BibitemShut
  {NoStop}%
\bibitem [{\citenamefont {Alba}\ \emph {et~al.}(2017)\citenamefont {Alba} \emph
  {et~al.}}]{Alba:2017mqu}%
  \BibitemOpen
  \bibfield  {author} {\bibinfo {author} {\bibfnamefont {P.}~\bibnamefont
  {Alba}} \emph {et~al.},\ }\href {\doibase 10.1103/PhysRevD.96.034517}
  {\bibfield  {journal} {\bibinfo  {journal} {Phys. Rev.}\ }\textbf {\bibinfo
  {volume} {D96}},\ \bibinfo {pages} {034517} (\bibinfo {year} {2017})},\
  \Eprint {http://arxiv.org/abs/1702.01113} {arXiv:1702.01113 [hep-lat]}
  \BibitemShut {NoStop}%
\bibitem [{\citenamefont {Ono}\ \emph {et~al.}(2019)\citenamefont {Ono} \emph
  {et~al.}}]{Ono:2019ndq}%
  \BibitemOpen
  \bibfield  {author} {\bibinfo {author} {\bibfnamefont {A.}~\bibnamefont
  {Ono}} \emph {et~al.},\ }\href {\doibase 10.1103/PhysRevC.100.044617}
  {\bibfield  {journal} {\bibinfo  {journal} {Phys. Rev. C}\ }\textbf {\bibinfo
  {volume} {100}},\ \bibinfo {pages} {044617} (\bibinfo {year} {2019})},\
  \Eprint {http://arxiv.org/abs/1904.02888} {arXiv:1904.02888 [nucl-th]}
  \BibitemShut {NoStop}%
\bibitem [{\citenamefont {Schenke}\ \emph {et~al.}(2010)\citenamefont
  {Schenke}, \citenamefont {Jeon},\ and\ \citenamefont
  {Gale}}]{Schenke:2010nt}%
  \BibitemOpen
  \bibfield  {author} {\bibinfo {author} {\bibfnamefont {B.}~\bibnamefont
  {Schenke}}, \bibinfo {author} {\bibfnamefont {S.}~\bibnamefont {Jeon}}, \
  and\ \bibinfo {author} {\bibfnamefont {C.}~\bibnamefont {Gale}},\ }\href
  {\doibase 10.1103/PhysRevC.82.014903} {\bibfield  {journal} {\bibinfo
  {journal} {Phys. Rev. C}\ }\textbf {\bibinfo {volume} {82}},\ \bibinfo
  {pages} {014903} (\bibinfo {year} {2010})},\ \Eprint
  {http://arxiv.org/abs/1004.1408} {arXiv:1004.1408 [hep-ph]} \BibitemShut
  {NoStop}%
\bibitem [{\citenamefont {Schenke}\ \emph {et~al.}(2011)\citenamefont
  {Schenke}, \citenamefont {Jeon},\ and\ \citenamefont
  {Gale}}]{Schenke:2010rr}%
  \BibitemOpen
  \bibfield  {author} {\bibinfo {author} {\bibfnamefont {B.}~\bibnamefont
  {Schenke}}, \bibinfo {author} {\bibfnamefont {S.}~\bibnamefont {Jeon}}, \
  and\ \bibinfo {author} {\bibfnamefont {C.}~\bibnamefont {Gale}},\ }\href
  {\doibase 10.1103/PhysRevLett.106.042301} {\bibfield  {journal} {\bibinfo
  {journal} {Phys. Rev. Lett.}\ }\textbf {\bibinfo {volume} {106}},\ \bibinfo
  {pages} {042301} (\bibinfo {year} {2011})},\ \Eprint
  {http://arxiv.org/abs/1009.3244} {arXiv:1009.3244 [hep-ph]} \BibitemShut
  {NoStop}%
\bibitem [{\citenamefont {Huovinen}\ and\ \citenamefont
  {Petreczky}(2010)}]{Huovinen:2009yb}%
  \BibitemOpen
  \bibfield  {author} {\bibinfo {author} {\bibfnamefont {P.}~\bibnamefont
  {Huovinen}}\ and\ \bibinfo {author} {\bibfnamefont {P.}~\bibnamefont
  {Petreczky}},\ }\href {\doibase 10.1016/j.nuclphysa.2010.02.015} {\bibfield
  {journal} {\bibinfo  {journal} {Nucl. Phys. A}\ }\textbf {\bibinfo {volume}
  {837}},\ \bibinfo {pages} {26} (\bibinfo {year} {2010})},\ \Eprint
  {http://arxiv.org/abs/0912.2541} {arXiv:0912.2541 [hep-ph]} \BibitemShut
  {NoStop}%
\bibitem [{\citenamefont {Bass}\ \emph {et~al.}(1998)\citenamefont {Bass} \emph
  {et~al.}}]{Bass:1998ca}%
  \BibitemOpen
  \bibfield  {author} {\bibinfo {author} {\bibfnamefont {S.}~\bibnamefont
  {Bass}} \emph {et~al.},\ }\href {\doibase 10.1016/S0146-6410(98)00058-1}
  {\bibfield  {journal} {\bibinfo  {journal} {Prog. Part. Nucl. Phys.}\
  }\textbf {\bibinfo {volume} {41}},\ \bibinfo {pages} {255} (\bibinfo {year}
  {1998})},\ \Eprint {http://arxiv.org/abs/nucl-th/9803035}
  {arXiv:nucl-th/9803035} \BibitemShut {NoStop}%
\bibitem [{\citenamefont {Bleicher}\ \emph {et~al.}(1999)\citenamefont
  {Bleicher} \emph {et~al.}}]{Bleicher:1999xi}%
  \BibitemOpen
  \bibfield  {author} {\bibinfo {author} {\bibfnamefont {M.}~\bibnamefont
  {Bleicher}} \emph {et~al.},\ }\href {\doibase 10.1088/0954-3899/25/9/308}
  {\bibfield  {journal} {\bibinfo  {journal} {J. Phys. G}\ }\textbf {\bibinfo
  {volume} {25}},\ \bibinfo {pages} {1859} (\bibinfo {year} {1999})},\ \Eprint
  {http://arxiv.org/abs/hep-ph/9909407} {arXiv:hep-ph/9909407} \BibitemShut
  {NoStop}%
\bibitem [{\citenamefont {Nunes~da Silva}\ \emph {et~al.}(2019)\citenamefont
  {Nunes~da Silva}, \citenamefont {Dobrigkeit~Chinellato}, \citenamefont
  {Derradi De~Souza}, \citenamefont {Hippert}, \citenamefont {Luzum},
  \citenamefont {Noronha},\ and\ \citenamefont
  {Takahashi}}]{NunesdaSilva:2018viu}%
  \BibitemOpen
  \bibfield  {author} {\bibinfo {author} {\bibfnamefont {T.}~\bibnamefont
  {Nunes~da Silva}}, \bibinfo {author} {\bibfnamefont {D.}~\bibnamefont
  {Dobrigkeit~Chinellato}}, \bibinfo {author} {\bibfnamefont {R.}~\bibnamefont
  {Derradi De~Souza}}, \bibinfo {author} {\bibfnamefont {M.}~\bibnamefont
  {Hippert}}, \bibinfo {author} {\bibfnamefont {M.}~\bibnamefont {Luzum}},
  \bibinfo {author} {\bibfnamefont {J.}~\bibnamefont {Noronha}}, \ and\
  \bibinfo {author} {\bibfnamefont {J.}~\bibnamefont {Takahashi}},\ }\href
  {\doibase 10.3390/proceedings2019010005} {\bibfield  {journal} {\bibinfo
  {journal} {MDPI Proc.}\ }\textbf {\bibinfo {volume} {10}},\ \bibinfo {pages}
  {5} (\bibinfo {year} {2019})},\ \Eprint {http://arxiv.org/abs/1811.05048}
  {arXiv:1811.05048 [nucl-th]} \BibitemShut {NoStop}%
\bibitem [{\citenamefont {Marrochio}\ \emph {et~al.}(2015)\citenamefont
  {Marrochio}, \citenamefont {Noronha}, \citenamefont {Denicol}, \citenamefont
  {Luzum}, \citenamefont {Jeon},\ and\ \citenamefont
  {Gale}}]{Marrochio:2013wla}%
  \BibitemOpen
  \bibfield  {author} {\bibinfo {author} {\bibfnamefont {H.}~\bibnamefont
  {Marrochio}}, \bibinfo {author} {\bibfnamefont {J.}~\bibnamefont {Noronha}},
  \bibinfo {author} {\bibfnamefont {G.~S.}\ \bibnamefont {Denicol}}, \bibinfo
  {author} {\bibfnamefont {M.}~\bibnamefont {Luzum}}, \bibinfo {author}
  {\bibfnamefont {S.}~\bibnamefont {Jeon}}, \ and\ \bibinfo {author}
  {\bibfnamefont {C.}~\bibnamefont {Gale}},\ }\href {\doibase
  10.1103/PhysRevC.91.014903} {\bibfield  {journal} {\bibinfo  {journal} {Phys.
  Rev. C}\ }\textbf {\bibinfo {volume} {91}},\ \bibinfo {pages} {014903}
  (\bibinfo {year} {2015})},\ \Eprint {http://arxiv.org/abs/1307.6130}
  {arXiv:1307.6130 [nucl-th]} \BibitemShut {NoStop}%
\bibitem [{\citenamefont {Denicol}\ \emph {et~al.}(2014)\citenamefont
  {Denicol}, \citenamefont {Jeon},\ and\ \citenamefont
  {Gale}}]{Denicol:2014vaa}%
  \BibitemOpen
  \bibfield  {author} {\bibinfo {author} {\bibfnamefont {G.}~\bibnamefont
  {Denicol}}, \bibinfo {author} {\bibfnamefont {S.}~\bibnamefont {Jeon}}, \
  and\ \bibinfo {author} {\bibfnamefont {C.}~\bibnamefont {Gale}},\ }\href
  {\doibase 10.1103/PhysRevC.90.024912} {\bibfield  {journal} {\bibinfo
  {journal} {Phys. Rev. C}\ }\textbf {\bibinfo {volume} {90}},\ \bibinfo
  {pages} {024912} (\bibinfo {year} {2014})},\ \Eprint
  {http://arxiv.org/abs/1403.0962} {arXiv:1403.0962 [nucl-th]} \BibitemShut
  {NoStop}%
\bibitem [{\citenamefont {Adam}\ \emph
  {et~al.}(2016{\natexlab{b}})\citenamefont {Adam} \emph
  {et~al.}}]{ALICE:2016kpq}%
  \BibitemOpen
  \bibfield  {author} {\bibinfo {author} {\bibfnamefont {J.}~\bibnamefont
  {Adam}} \emph {et~al.} (\bibinfo {collaboration} {ALICE}),\ }\href {\doibase
  10.1103/PhysRevLett.117.182301} {\bibfield  {journal} {\bibinfo  {journal}
  {Phys. Rev. Lett.}\ }\textbf {\bibinfo {volume} {117}},\ \bibinfo {pages}
  {182301} (\bibinfo {year} {2016}{\natexlab{b}})},\ \Eprint
  {http://arxiv.org/abs/1604.07663} {arXiv:1604.07663 [nucl-ex]} \BibitemShut
  {NoStop}%
\bibitem [{\citenamefont {Giacalone}\ \emph {et~al.}(2016)\citenamefont
  {Giacalone}, \citenamefont {Yan}, \citenamefont {Noronha-Hostler},\ and\
  \citenamefont {Ollitrault}}]{Giacalone:2016afq}%
  \BibitemOpen
  \bibfield  {author} {\bibinfo {author} {\bibfnamefont {G.}~\bibnamefont
  {Giacalone}}, \bibinfo {author} {\bibfnamefont {L.}~\bibnamefont {Yan}},
  \bibinfo {author} {\bibfnamefont {J.}~\bibnamefont {Noronha-Hostler}}, \ and\
  \bibinfo {author} {\bibfnamefont {J.-Y.}\ \bibnamefont {Ollitrault}},\ }\href
  {\doibase 10.1103/PhysRevC.94.014906} {\bibfield  {journal} {\bibinfo
  {journal} {Phys. Rev. C}\ }\textbf {\bibinfo {volume} {94}},\ \bibinfo
  {pages} {014906} (\bibinfo {year} {2016})},\ \Eprint
  {http://arxiv.org/abs/1605.08303} {arXiv:1605.08303 [nucl-th]} \BibitemShut
  {NoStop}%
\end{thebibliography}%

\end{document}